\documentclass[5p, times,authoryear]{elsarticle}

\usepackage[dvipsnames]{xcolor}
\usepackage[utf8]{inputenc}
\usepackage{amsmath}
\usepackage{hyperref}
\usepackage{graphicx}
\usepackage{siunitx}
\usepackage{array}
\usepackage[ruled,vlined]{algorithm2e}
\usepackage{soulutf8}
\usepackage{tikz}
\usetikzlibrary{decorations.pathreplacing}
\usepackage{float}
\usepackage{balance}
\usepackage{amssymb}
\usepackage{url}
\usepackage{xspace}
\usepackage{tcolorbox}
\usepackage{enumitem}
\usepackage{multirow}
\usetikzlibrary{shapes}
\usetikzlibrary{patterns, fit}
\usepackage{adjustbox}
\usepackage[T1]{fontenc}
\usepackage{pifont}
\usepackage{mathtools}
\usepackage[super]{nth}
\usepackage{mdframed}
\usepackage{xfrac}
\usepackage{natbib}
\setcitestyle{aysep={}}
\usepackage[section]{placeins}
\usepackage{afterpage}
\usepackage{needspace}
\usepackage{afterpage}

\DeclarePairedDelimiter\ceil{\lceil}{\rceil}

\makeatletter
\g@addto@macro{\UrlBreaks}{\UrlOrds}

\newcommand{\highlight}[1]{\begin{tcolorbox}[leftrule=1mm,rightrule=1mm,toprule=0mm,bottomrule=0mm,left=2pt,right=2pt,top=1pt,bottom=1pt]
#1
\end{tcolorbox}
}

\newcounter{findings}
\newcommand{\finding}[2]{\begin{mdframed}{\textbf{#1 \thefindings}: #2}\end{mdframed} \stepcounter{findings}}

\newcounter{customdefinitions}
\newcommand{\customdefinition}[2]{\begin{mdframed}{\textbf{#1 \thecustomdefinitions}: #2}\end{mdframed} \stepcounter{customdefinitions}}

\newboolean{showcomments}
\setboolean{showcomments}{true}
\ifthenelse{\boolean{showcomments}}
 { \newcommand{\mynote}[2]{
      \fbox{\bfseries\sffamily\scriptsize#1}
        {\small$\blacktriangleright$\textsf{\emph{#2}}$\blacktriangleleft$}}}
        { \newcommand{\mynote}[2]{}}

\newcommand{\obfuscapk}[0]{\textsc{Obfuscapk}\xspace}

\makeatletter
\newcommand{\ostar}{\mathbin{\mathpalette\make@circled\star}}
\newcommand{\make@circled}[2]{%
  \ooalign{$\m@th#1\smallbigcirc{#1}$\cr\hidewidth$\m@th#1#2$\hidewidth\cr}%
}
\newcommand{\smallbigcirc}[1]{%
  \vcenter{\hbox{\scalebox{0.97778}{$\m@th#1\bigcirc$}}}%
}
\makeatother

\newcolumntype{g}{>{\columncolor{gray!30}}l}
\newcolumntype{h}{>{\columncolor{gray!30}}c}

\newcommand{\approachname}[0]{\textsc{DexRay}\xspace}
\newcommand{\drebin}[0]{\textsc{Drebin}\xspace}
\newcommand{\az}[0]{\textsc{AndroZoo}\xspace}
\newcommand{\gp}[0]{Google Play\xspace}
\newcommand{\rd}[0]{\textsc{R2-D2}\xspace}
\newcommand{\yx}[0]{Ding et al.\xspace}

\newcolumntype{P}[1]{>{\centering\arraybackslash}p{#1}}

\begin{document}
\begin{frontmatter}

\title{\approachname: A Simple, yet Effective Deep Learning Approach to Android Malware Detection based on Image Representation of Bytecode}

\author[inst1,inst2]{Nadia Daoudi\corref{cor1}}
\ead{nadia.daoudi@list.lu}

\author[inst2]{Jordan Samhi}
\ead{jordan.samhi@uni.lu}

\author[inst2]{Abdoul Kader Kaboré}
\ead{abdoulkader.kabore@uni.lu}

\author[inst3]{Kevin Allix}
\ead{kevin.allix@ext.uni.lu}

\author[inst2]{Tegawendé F. Bissyandé}
\ead{tegawende.bissyande@uni.lu}

\author[inst2]{Jacques Klein}
\ead{jacques.klein@uni.lu}

\cortext[cor1]{Corresponding author}

\address[inst1]{Luxembourg Institute of Science and Technology.}
\address[inst2]{SnT, University of Luxembourg.}
\address[inst3]{Independent Researcher}

\begin{abstract}
Computer vision has witnessed several advances in recent years, with unprecedented performance provided by deep representation learning research. 
Image formats thus appear attractive to other fields such as malware detection, where deep learning on images alleviates the need for comprehensive hand-crafted features generalising to different malware variants.
We contribute with a first building block by developing and assessing a baseline pipeline for image-based malware detection with straightforward steps.
We propose \approachname, which converts the bytecode of the app DEX files into grey-scale ``vector'' images and feeds them to a 1-dimensional Convolutional Neural Network model.
We view \approachname as foundational due to the exceedingly basic nature of the design choices, allowing to infer what could be a minimal performance that can be obtained with image-based learning in malware detection.
The performance of \approachname evaluated on over 158k apps 
demonstrates that our approach is effective with a high detection rate (F1-score $=0.96$).
This paper contributes to the domain of Deep Learning based Malware detection by providing a sound, simple, yet effective approach (with available artefacts) that can be the basis to scope the many profound questions that will need to be investigated to fully develop this domain.

\end{abstract}
\begin{keyword}
Android \sep  Security \sep Malware detection \sep Machine learning
\end{keyword}
\end{frontmatter} 







\section{Introduction}\label{sec:introduction}

Automating malware detection is a key concern in the research and practice communities. 
There is indeed a huge number of samples to assess, making it challenging to consider any manual solutions.
Consequently, several approaches have been proposed in the literature to automatically detect malware~\citep{kang2015detecting,10.1145/2592791.2592796, 6680837, AndroSimilar}.
However, current approaches remain limited and detecting all malware is still considered  an unattainable dream.
A recent report from McAfee~\citep{mcafeeReport} shows that mobile malware has increased by 15\% between the first and the second quarter of 2020.
Moreover, Antivirus companies and Google, the official Android market maintainer, have disclosed that malware
apps become more and more sophisticated and threaten users' privacy and security~\citep{googleReport,malwarebytesReport,kasperskyReport}. 

To prevent the spread of malware and help security analysts, researchers have leveraged Machine Learning approaches~\citep{drebin, revealdroid, mamadroid, ANASTASIA, droidcat, DroidDolphin, BRIDEMAID} that rely on features extracted statically, dynamically, or in an hybrid manner.
While a dynamic approach extracts the features when the apps are running, a static approach relies 
exclusively on the artefacts present in the APK file.
As for hybrid approaches, they combine both statically and dynamically retrieved features.
Both static and dynamic approaches require manual engineering of the features in order to select some information that can, to some extent, approximate the behaviour of the apps. 
Also, the hand-crafted features might miss some information that is relevant to distinguish malware from benign apps. 
Unfortunately, good feature engineering remains an open problem in the literature due to the challenging task of characterising maliciousness in terms of app artefacts.

Recently, a new wave of research around representation of programs for malware detection has been triggered. 
Microsoft and Intel Labs have proposed STAMINA\footnote{\url{https://www.microsoft.com/security/blog/2020/05/08/microsoft-researchers-work-with-intel-labs-to-explore-new-deep-learning-approaches-for-malware-classification/}}, a Deep Learning approach that detects malware based on image representation of binaries. 
This approach is built on deep transfer learning from computer vision, and it was shown to be highly effective in detecting malware.
In the Android research community, image-based malware detection seems to be attractive. 
Some approaches have investigated the image representation of the APK's code along with deep learning techniques.
However, they have directly jumped to non-trivial representations (e.g., colour) and/or leveraged complex learning architectures (e.g., Inception-v3~\citep{inceptionv3}).
Furthermore, some supplementary processing steps are applied which may have some effects on the performance yielded.
Besides, they create square or rectangular images, which might distort the bytecode sequences from the DEX files, therefore at the same time losing existing \emph{locality} and creating artificial \emph{locality}. 
Indeed, given that the succession of pixels in the image depends on the series of bytes in the DEX files, converting the bytecode to a ``rectangular'' image can result in having patterns that start at the end of a line (i.e., row of the image) and finish at the beginning of the next line in the image.
Moreover, related approaches leverage 2-dimensional convolutional neural networks, which perform convolution operations that take into consideration the pixels in the 2-d neighbourhood using 2-d kernels.
Since there is no relationship between pixels belonging to the same column (i.e., pixels that are above/below each others) of a ``rectangular'' image representation of code, the use of 2-d convolutional neural networks does not sound appropriate.

Image-based representation is, still, a sweet spot for leveraging advances in deep learning as it 
has been demonstrated with the impressive results reported in the computer vision field.
To ensure the development of the research direction within the community, we propose to investigate a straightforward ``vector'' image-based malware detection approach leveraging both a simple image representation and a basic neural network architecture. 
Our aim is to deliver a foundational framework towards enabling the community to build novel state-of-the-art approaches in malware detection. 
This paper presents the initial insights into the performance of a baseline that is made publicly available to the community.

This paper, which is an extension of our prior work, makes the following contributions:
\begin{itemize}
    \item We propose \approachname, a foundational image-based Android malware detection approach that converts the sequence of raw byte code contained in the DEX file(s) into a simple grey-scale ``vector'' image of size (1, 128*128).
    Features extraction and classification are assigned to a 1-dimensional-based convolutional neural network model.
    \item We evaluate \approachname on a large dataset of more than 158k malware and goodware apps.
    \item We discuss the performance achievement of \approachname in comparison with prior work, notably the state-of-the-art \drebin approach as well as related work leveraging APK to image representations.
    \item We evaluate the performance of \approachname on detecting new malware apps.
    \item We investigate the impact of image-resizing on the performance of \approachname.
    \item We assess the resilience of \approachname and \drebin to apps' obfuscation.
    \item We make the source code of \approachname publicly available along with the dataset of apps and images.
\end{itemize}
Additionally, compared to our previous work, we propose a novel investigation into two aspects of \approachname:
\begin{itemize}
    \item We evaluate the performance of \approachname on malware family classification.
    \item We study the possibility to localise malicious code in image representation of Android apps.
\end{itemize}

\section{Approach}
\label{sec:approach}

In this section, we present the main basic blocks of \approachname that are illustrated in Fig.~\ref{fig:worflow}.
We present in Section~\ref{sec:approach:images} our image representation process which covers steps (1.1) and (1.2) in Fig.~\ref{fig:worflow}. As for step (2), it is detailed in Section~\ref{sec:approach:deeplearning}

\begin{figure*}[!htbp]
    \centering
    \input{tikz/overview}
    \caption{\approachname basic building blocks.}
    \label{fig:worflow}
\end{figure*}

\subsection{Image representation of Android apps}
\label{sec:approach:images}

Android apps are delivered in the form of packages called APK (Android Package) whose  size can easily reach several MegaBytes~\citep{androzoo}. 
The APK includes the bytecode (DEX files), some resource files, native libraries, and other assets.

\textbf{(1.1) Bytecode extraction:} Our approach focuses on code, notably the applications' bytecode, i.e., DEX files, where the app behaviour is supposed to be implemented. 

\begin{figure*}[!htbp]
    \centering
    \input{tikz/images_generation}
    \caption{Process of image generation from dalvik bytecode. \ding{182}: bytecode bytes' vectorisation; \ding{183}: Mapping bytes to pixels.}
    \label{fig:image_generation}
\end{figure*}

\textbf{(1.2) Conversion into image:} Our straightforward process for converting DEX files into images is presented in Fig.~\ref{fig:image_generation}.
We concatenate all the DEX files present in the APK as a single byte stream vector (step \ding{182} in Fig.~\ref{fig:image_generation}).
This vector is then converted to a grey-scale ``vector'' image by considering each byte as an 8-bit pixel (step \ding{183} in Fig.~\ref{fig:image_generation}).
Given that apps can widely differ in their code size, the size of the resulting ``vector'' images is also different. 
Note that the size of our ``vector'' image representation refers to the width of the image, since its height is 1.
To comply with the constraints~\citep{LeCun2015} of off-the-shelf deep learning architectures for images, we leverage a standard image resizing algorithm\footnote{\url{https://www.tensorflow.org/api_docs/python/tf/image/resize}} to resize our ``vector'' images to a fixed-size.
For our experiments, we have selected a size of (1, 128x128). 
We also investigate the impact of resizing on the performance of \approachname in Section~\ref{sec:evaluation:resizing}.

We remind that related image-based Android malware detectors (e.g., \citep{R2D2,ding2020android}) leverage a ``rectangular'' image representation, which might destroy the succession of bytes in the DEX files (i.e., the succession of pixels in the rectangular image). 
Moreover, this representation usually requires padding to have a rectangular form of the image.
The complete procedure that shows the difference between our ``vector'' image and related approaches ``rectangular'' image generation is detailed in Algorithm~\ref{algo:image_vector_generation} and Algorithm~\ref{algo:image_square_generation} respectively.
While there is more advanced ``rectangular'' image representations, Algorithm~\ref{algo:image_square_generation} shows an illustration with a basic ``square'' grey-scale image of size (128, 128).

\medskip

\begin{algorithm}[!ht]
\SetAlgoLined
\textbf{Input:} APK file \\
\textbf{Output:} 8-bit grey-scale ``vector'' image of size (1, 128x128) \\
\smallskip
 bytestream $\gets \emptyset$ \\
 \For{dexFile in APK}{
    bytestream $\gets$ bytestream + dexFile.toByteStream()
 }
 l $\gets$ bytestream.length() \\

 img $\gets$ generate8bitGreyScaleVectorImage(bytestream, l) \\
 img.resize\_to\_size(height=1, width=128x128) \\
 img.save()
 \caption{Algorithm describing 8-bit grey-scale ``vector'' image generation from an APK}
 \label{algo:image_vector_generation}
\end{algorithm}

\medskip

\medskip

\begin{algorithm}[!ht]
\SetAlgoLined
\textbf{Input:} APK file \\
\textbf{Output:} 8-bit grey-scale ``square'' image of size (128, 128) \\
\smallskip
 bytestream $\gets \emptyset$ \\
 \For{dexFile in APK}{
    bytestream $\gets$ bytestream + dexFile.toByteStream()
 }
 l $\gets$ bytestream.length() \\
 sqrt $\gets \ceil{\sqrt{l}}$ \\
 sq $\gets \text{sqrt}^2$ \\
 \While{bytestream.length() $\neq$ sq}{
    bytestream $\gets$ bytestream + "$\setminus$x00" // padding with zeros\\
 }
 // At this point, bytestream is divided in $sqrt$ part of length $sqrt$ \\
 // In other words, it is represented as a $sqrt \times sqrt$ matrix \\
 
 img $\gets$ generate8bitGreyScaleSquareImage(bytestream, sqrt) \\
 img.resize\_to\_size(height=128, width=128) \\
 img.save()
 \caption{Algorithm describing 8-bit grey-scale ``square'' image generation from an APK}
 \label{algo:image_square_generation}
\end{algorithm}

\medskip
Comparing the two Algorithms, we can notice that our image representation is very basic since it uses the raw bytecode ``as it is'', without any segmentation or padding needed to achieve a ``square'' or ``rectangular'' form of the image. 
In the remainder of this paper, we use ``image'' to refer to our ``vector'' image representation of code.

\subsection{Deep Learning Architecture}
\label{sec:approach:deeplearning}
Convolutional Neural Networks (CNN) are specific architectures for deep learning. They have been particularly successful for learning tasks involving images as inputs. 
CNNs learn representations by including three types of layers: Convolution, Pooling, and Fully connected layers~\citep{yamashita2018convolutional}.
We describe these notions in the following:

\begin{figure*}[!htbp]
    \centering
   \resizebox{0.8\linewidth}{!}{
    \input{tikz/architecture_vect}}
    \caption{Our Convolutional Neural Network architecture}
    \label{fig:architecture}
\end{figure*}

\begin{itemize}
	\item The convolution layer is the primary building block of convolutional neural networks.
	This layer extracts and learns relevant features from the input images. 
	Specifically, a convolutional layer defines a number of filters (matrices) that detect patterns in the image.
	Each filter is convolved across the input image by calculating a dot product between the filter and the input.
	The filters' parameters are updated and learned during the network's training with the backpropagation algorithm~\citep{8029130,8286426}.
	The convolution operation creates feature maps that pass first through an activation function, and then they are 
	received as inputs by the next layer to perform further calculation.
	\item A pooling layer is generally used after a convolutional layer. 
	The aim of this layer is to downsize the feature maps, so the number of parameters and the time of computation is reduced~\citep{ke2018computer}.
	Max pooling and Average pooling are two commonly used methods to reduce the size of the feature maps received by a pooling layer in CNNs~\citep{MixedPooling}.
	\item In the Fully connected layer, each neuron is connected to all the neurons in the previous and the next layer. 
	The feature maps from the previous convolution/pooling layer are flattened and passed to this layer in order to make the classification decision. 
\end{itemize}

Among the variety of Deep-Learning architectures presented in the literature, CNNs constitute a strong basis for deep learning with images. We further propose to keep a minimal configuration of the presented architecture by implementing a convolutional neural network model that makes use of 1-dimensional convolutional layers. 
In this type of layers, the filter is convolved across one dimension only, which reduces the number of parameters and the time of the training.
Also, 1-d convolutional layers are the best suited for image representation of code since the pixels represent the succession of bytecode bytes' from the apps.
The use of 1-d convolution on our images can be thought of as sliding a convolution window over the sequences of bytes searching for patterns that can distinguish malware and benign apps.

We present in Fig.~\ref{fig:architecture} our proposed architecture, which contains two 1-dimensional convolutional/pooling layers that represent the extraction units of our neural network. 
The feature maps generated by the second max-pooling layer are then flattened and passed to a dense layer that learns to discriminate malware from benign images. 
The second dense layer outputs the detection decision.

\section{Study Design}
In this section, we first overview the research questions that we investigate.  
Then, we present the datasets and the experimental setup used to answer these research questions.

\subsection{Research Questions}
In this paper, we consider the following six main research questions:

\begin{itemize}
    \item \textbf{RQ1:} How effective is \approachname in the detection of Android malware?
    \item \textbf{RQ2:} How effective is \approachname in detecting new Android malware?
    \item \textbf{RQ3:} What is the impact of image-resizing on the performance of \approachname?
    \item \textbf{RQ4:} How does app obfuscation affect the performance of \approachname?
    \item \textbf{RQ5:} How effective is \approachname in the classification of Android malware families?
    \item \textbf{RQ6:} What can \approachname tell us about the possibility to localise malicious code?
\end{itemize}

\subsection{Dataset}\label{sec:eval:dataset}

\textbf{Initial Dataset.}
To conduct our experiments, we collect malware and benign apps from \az~\citep{androzoo} which 
contains, at the time of writing, more than 16 million Android apps.
\az crawls apps from several sources, including the official Android market \gp\footnote{\url{https://play.google.com/store}}.
We have collected \az apps that have their compilation dates between January 2019 and May 2020. Specifically, our dataset contains 134\,134 benign, and 71\,194 malware apps. 
Benign apps are defined as the apps that have not been detected by any antivirus from 
VirusTotal\footnote{\url{https://www.virustotal.com/}}. 
The malware collection contains the apps that have been detected by at least two antivirus engines, similarly to what is done in \drebin{}~\citep{drebin}.

\noindent
\textbf{Obfuscation process.} 
In our experiments, we further explore the impact of app obfuscation.
Therefore, we propose to generate obfuscated apps, by relying on the state-of-the-art Android app obfuscator \obfuscapk\footnote{\url{https://github.com/ClaudiuGeorgiu/Obfuscapk}}~\citep{aonzo2020obfuscapk}.
This tool takes an Android app (APK) as input and outputs a new APK with its bytecode obfuscated.
The obfuscation to apply can be configured by selecting a combination of more than 20 obfuscation processes.
These obfuscation processes range from light obfuscation (e.g., inserting \texttt{nop} instructions, removing debug information, etc.) to heavyweight obfuscation (e.g., replacing method calls to reflection calls, strings encryption, etc.).

To evaluate our approach's resiliency to obfuscation, we decided to use seven different obfuscation processes: 
(1) renaming classes, (2) renaming fields, (3) renaming methods, (4) encrypting strings, (5) overloading methods, (6) adding call indirection, (7) transforming method calls with reflection\footnote{\url{https://www.oracle.com/technical-resources/articles/java/javareflection.html}}.
Thus, the resulting APK is considered to be highly obfuscated.
Since \obfuscapk is prone to crash, we were not able to get the obfuscated version for some apps. 

We generate the images for the non-obfuscated dataset (the apps downloaded from \az), and for the obfuscated samples.
The image generation process is detailed in Section~\ref{sec:approach:images}.

Since we compare our method with \drebin's approach, we have also extracted \drebin's features from
the exact same apps we use to evaluate our approach.
In our experiments, we only consider the apps from the non-obfuscated and the obfuscated datasets for which we have (a) successfully generated their images, and (b) successfully extracted their features for \drebin.
Consequently, our final dataset contains \num{61809} malware and \num{96994} benign apps.
We present in Table~\ref{tab:dataset} a summary of our dataset.

\begin{table}[!ht]
	\begin{center}
		\caption{Dataset summary}
		\label{tab:dataset}
		\begin{tabular}{ |P{5cm} | P{1.3cm} | P{1cm}| }
		\hline
			&  \textbf{malware apps}  & \textbf{benign apps}  \\
			\hline
			\textbf{Initial Set} & 71\,194 & 134\,134  \\
			\textbf{Removed because of Image generation or obfuscation failure} & \num{9027} &  \num{36652} \\
			
			\textbf{Removed because of \drebin extraction failure} & 358 & 488  \\ \hline
			\textbf{Final Set} &  61\,809 & 96\,994 \\ \hline
		\end{tabular}
	\end{center}
\end{table}

\subsection{Empirical Setup}\label{sec:empr:setup}
\textbf{Experimental validation.} 
We evaluate the performance of \approachname using the Hold-out technique~\citep{raschka2018model}.
Specifically, we shuffle our dataset and split it into 80\% training, 10\% validation, and 
10\% test.
We train our model on the training dataset, and we use the apps in the validation set to tune the 
hyper-parameters of the network.
After the model is trained, we evaluate its performance using the apps in the test set.
This process is repeated ten times by shuffling the dataset each time and splitting it into training, validation, and test sets. 
We repeat the Hold-out technique in order to verify that our results do not depend on a specific 
split of the dataset.

We set to 200 the maximum number of epochs to train the network, and we stop the training process if 
the accuracy score on the validation dataset does not improve for 50 consecutive epochs.
As for the models' parameters, we use \texttt{kernel\_size=12}, and \texttt{activation=relu} for the two
convolution layers, and we set their number of filters to 64 and 128 respectively.
We also use \texttt{pool\_size=12} for the two max-pooling layers, and  \\ \texttt{activation=sigmoid} for the two dense layers. The number of neurons in the first dense layer is set to 64. 
As for the output layer, it contains one neuron that predicts the class of the apps.
We rely on four performance measures to assess the effectiveness of \approachname: Accuracy, Precision, Recall, and F1-score.
Our experiments are conducted using the TensorFlow library\footnote{\url{https://www.tensorflow.org}}.

\noindent
\textbf{State-of-the-art approaches to compare with.}
We present in this section the three Android malware detectors used in our experimental comparison: \drebin, \rd, and \yx.
While there are many ML-based Android malware approaches, only few are available~\citep{LessonsLearnt}.
We have selected \drebin since it is among the highly performing detectors that have their implementation available.
Regarding image-based Android malware detectors, \rd and \yx are the two approaches that are the most similar to \approachname.

{\itshape \drebin~\citep{drebin}: }
We compare \approachname against \drebin, the state-of-the-art approach in machine learning based malware detection. 
\drebin extracts features that belong to eight categories: Hardware components, Requested permissions, App 
components, Filtered intents, Restricted API calls, Used permissions, Suspicious API calls, and Network addresses.
These features are extracted from the disassembled bytecode and the Manifest file.
The combination of extracted features is used to create an $n$-dimensional vector space where $n$ is 
the total number of extracted features.
For each app in the dataset, an n-dimensional binary vector space is created. 
A value of 1 in the vector space indicates that the feature is present in the app. 
If a feature does not exist in the app, its value is set to 0.
\drebin feeds the vectors space to a Linear SVM classifier, and it uses the trained model to predict if an 
app is malware or goodware.
In this study, we use a replicated version~\citep{LessonsLearnt} of \drebin that we run on our dataset.

{\itshape \rd~\citep{R2D2}} is an Android malware detector that converts the bytecode of an APK file (i.e., DEX files) into RGB colour images.
Specifically, the hexadecimal from the bytecode is translated to RGB colour.
\rd is a CNN-based approach that trains Inception-v3~\citep{inceptionv3} with the coloured images to
predict malware.
In their paper, the authors state that their approach is trained with more than 1.5 million samples.
However, they do not clearly explicit how many apps are used in the test nor the size fixed for the coloured images.
The authors provide a link to the materials of their experiment\footnote{\url{http://r2d2.twman.org}},
but we could not find the image generator nor the original apps used to evaluate their approach in
Section IV-C of their paper.
Only the generated images are available.
As a result, we were unable to reproduce their experiment to compare with our model.
Instead, we compare directly with the results reported in their paper.

{\itshape \yx~\citep{ding2020android}} proposes to convert the DEX files into (512, 512) grey-scale images in order to feed them to a deep 
learning model.
The authors experiment with two CNN-based models: The first one, we note \texttt{Model1}, contains four 
convolutional layers, four pooling layers, a fully-connected hidden layer, and a fully-connected output layer.
\texttt{Model2}, which is the second model, has the same architecture as \texttt{Model1} but with an 
additional high-order feature layer that is added just after the first pooling layer.
Basically, this layer contains the multiplication of each two adjacent feature maps from the 
previous layer, as well as the feature maps themselves.
Since neither the dataset nor the implementation of \yx's models is publicly available, we also 
rely on the results reported in \yx's manuscript.

\section{Study Results}
\label{sec:evaluation}

\subsection{\textbf{RQ1:} How effective is \approachname in the detection of Android malware?}~\label{sec:evaluation_rq1}

In this section, we assess the performance of \approachname on Android malware detection. 
We consider the performance against a ground truth dataset (the non-obfuscated apps introduced in Section~\ref{sec:eval:dataset}) as well as a comparison against prior approaches.
We rely on the experimental setup presented in Section~\ref{sec:empr:setup} and we test the performance of \approachname on \num{15880} apps (10\% of the non-obfuscated apps).

We report in Table~\ref{tab:RQ1_and_2_scores} the scores of the ten training/test of \approachname as the average of Accuracy, Precision, Recall, and F1-score.

\begin{table}[!ht]
	\begin{center}
		\caption{Performance of \approachname against \drebin on our experimental dataset}
		\label{tab:RQ1_and_2_scores}
		\begin{tabular}{|c | c | c | c | c |}
		\hline
			&  \textbf{Accuracy}  & \textbf{Precision}& \textbf{Recall} & \textbf{F1-score} \\
			\hline
			\textbf{\approachname} & 0.97  & 0.97 & 0.95 & 0.96\\ 
			\hline
			\textbf{\drebin} & 0.97  & 0.97 & 0.94 & 0.96\\ 
 			\hline
		\end{tabular}
	\end{center}
\end{table} 

Overall, as shown in Table~\ref{tab:RQ1_and_2_scores}, \approachname reaches an average score of 0.97, 0.97, 0.95, and 0.96 for Accuracy, Precision, Recall, and F1-score respectively. 
The reported results show the high effectiveness of \approachname in detecting Android malware.  
We further compare the performance of \approachname against three Android malware detectors in the following.

\subsubsection{ Comparison with \drebin.}
To assess the effectiveness of \approachname, we compare our results against a state-of-the-art Android malware detector that relies on static analysis: \drebin.
Specifically, we evaluate \drebin using the same exact non-obfuscated apps we use to evaluate
\approachname, and the same experimental setup described in Section~\ref{sec:empr:setup}. 
Moreover, we use the same split of the dataset for the Hold-out technique to evaluate the two approaches.
We report the average of Accuracy, Precision, Recall, and F1-score of \drebin's evaluation in 
Table~\ref{tab:RQ1_and_2_scores}.

We notice that \drebin and \approachname achieve the same Accuracy, Precision, and the F1-score. 
As for the Recall \approachname slightly outperforms \drebin with a difference of 0.01.

\subsubsection{Comparison against other Image-based malware detection Approaches.}\label{eval:comp:image}

\begin{table}[!ht]
	\begin{center}
		\caption{Authors-reported performance of \rd and \yx on different and less-significant datasets}
		\label{tab:RQ1_and_r2d2}
            \resizebox{\linewidth}{!}{
		\begin{tabular}{|c| c| c| c| c| }
		\hline
			&  \textbf{Accuracy}  & \textbf{Precision}& \textbf{Recall} & \textbf{F1-score} \\
			\hline
			\textbf{\approachname} & 0.97  & 0.97 & 0.95 & 0.96\\ 
			\hline
			\textbf{\rd} & 0.97  & 0.96 & 0.97 & 0.97\\
			\hline
			\textbf{\yx-Model~1} & 0.94   & - & 0.93 & - \\
			\hline
			\textbf{\yx-Model~2} & 0.95  & - & 0.94 & -\\ 
            \hline
		\end{tabular}}
	\end{center}
\end{table}

We present in Table~\ref{tab:RQ1_and_r2d2} the detection performance of \rd and \yx's approaches as reported in their original publications.
We note that \approachname and these two approaches are not evaluated using the same experimental setup and dataset. 
Specifically, \rd is trained on a huge collection of 1.5 million apps, but it is evaluated on a small 
collection of 5482 images.
In our experiments, we have conducted an evaluation on \num{15880} test apps.
We note that we have inferred the size of the test set based on \rd publicly available images.
Also, the scores we report for \approachname are the average of the scores achieved by ten different 
classifiers, each of which is evaluated on different test samples.
\rd scores are the results of a single train/test experiment which makes it difficult to properly compare the two approaches.

Similarly, \yx's experiments are conducted using the cross-validation technique, and both \texttt{Model1} and \texttt{Model2} are trained and evaluated using a small dataset of 4962 samples.
Overall, we note that in terms of Recall, Precision and Accuracy, \approachname achieves performance metrics that are on par with prior work. 

\highlight{
\textbf{RQ1 answer:} \approachname is a straightforward approach to malware detection which yields performance metrics that are comparable to the state of the art. Furthermore, it demonstrates that its simplicity in image generation and network architecture has not hindered its performance when compared to similar works presenting sophisticated configurations}

\subsection{\textbf{RQ2:} How effective is \approachname in detecting new Android malware?}
\label{sec:evaluation:time} 

Time decay~\citep{tesseract} and model ageing~\citep{8806731} refer to the situation when the performance of ML classifiers drops after they are tested on new samples. Previous works have shown that this situation is common with state of the art approaches~\citep{10.1145/3372297.3417291,Allix2016:emse}.
In this section, we aim to assess how does model ageing affect the performance of \approachname. 
Specifically, we investigate if \approachname can detect new malware when all the samples in its training set are older than the samples in its test set---a setting that Tesseract authors called \emph{Temporally consistent}~\citep{tesseract}.
To this end, we split our dataset into two parts based on the date specified in the DEX file of the apps. 
The apps from 2019 are used to train and tune the hyper-parameters of the model, and the apps from 2020 are used for the test. 
The training and validation datasets consist of \num{113951} and \num{28488} apps respectively (i.e., 80\% and 20\% of 2019 dataset). 
As for the test dataset, it contains \num{16364} malware and benign apps from 2020.
We report our results on Table~\ref{tab:eval_time}. 

\begin{table}[!ht]
	\begin{center}
		\caption{Impact of model ageing on the performance of \approachname}
		\label{tab:eval_time}
           \resizebox{\linewidth}{!}{
		\begin{tabular}{ |P{3.2cm} | c |c |c |c| }
		\hline
			 & \textbf{Accuracy}  & \textbf{Precision}& \textbf{Recall} & \textbf{F1-score} \\
			\hline
			
			\textbf{\approachname results from RQ1} & 0.97  & 0.97 & 0.95 & 0.96\\ 
			  \hline
			  \textbf{\approachname (Temporally Consistent)} & 0.97 & 0.97 & 0.98 & 0.98\\
			 \hline
		\end{tabular}}
	\end{center}
\end{table} 

We notice that \approachname detects new malware apps with a high detection rate. 
Specifically, it achieves detection scores of 0.97, 0.97, 0.98, 0.98 for Accuracy, Precision, Recall, and F1-score respectively.
Compared to its effectiveness reported in RQ1 in Section~\ref{sec:evaluation_rq1}, we notice that \approachname has reported higher Recall in this {\em Temporally consistent} experiment. 
This result could be explained by the composition of the malware in the training and test set. 
Indeed, malicious patterns in the test apps have been learned during the training that contains a representative set of Android malware from January to December 2019.
The high performance of \approachname on new Android apps demonstrates its ability to generalise and its robustness against model ageing.

\begin{figure}[!htb]
    \centering
    \includegraphics[width=\linewidth]{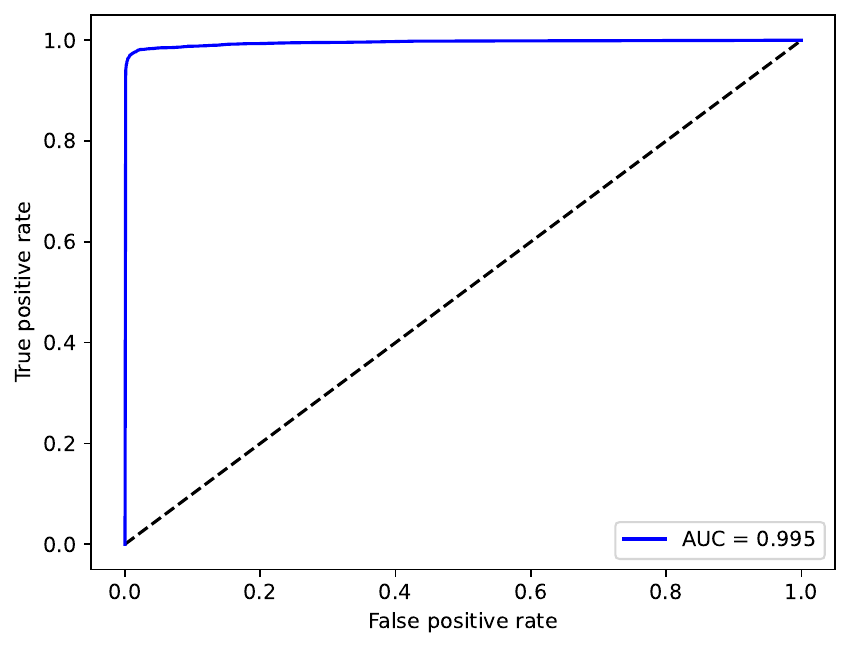}
    \caption{ROC curve of \approachname evaluated against time decay}
    \label{fig:roc_dexray_drebin}
\end{figure}

In this experiment, it is not possible to use the 10-times Hold-out technique because we only have one model trained on apps from 2019 and tested on new apps from 2020.
To conduct an in-depth evaluation of the detection capabilities of this model, we also examine the receiver operating characteristic (ROC) curve. 
The ROC curve shows the impact of varying the threshold of the positive class on the performance of the classifier. It can be generated by plotting the true positive rate (TPR) against the false positive rate (FPR).
We present in Figure~\ref{fig:roc_dexray_drebin} the ROC curve of our model.

The ROC curve of \approachname further confirms its high effectiveness. 
As the area under the curve (AUC) can have a maximum value of 1, our approach has reached a very high AUC of 0.995. 

It is noteworthy to remind that Tesseract~\citep{tesseract} reported that \drebin performance was very significantly (negatively) impacted when tested in a \emph{Temporally Consistent} setting.

\highlight{
\textbf{RQ2 answer:} 
\approachname's high performance is maintained in a \emph{Temporally Consistent} evaluation. 
When tested on new Android apps, our approach has achieved very high detection performance, with an AUC of 0.995.
}
\subsection{\textbf{RQ3:} What is the impact of image-resizing on the performance of \approachname?}
\label{sec:evaluation:resizing}

\begin{figure}[!htb]
    \centering
    \includegraphics[width=\linewidth]{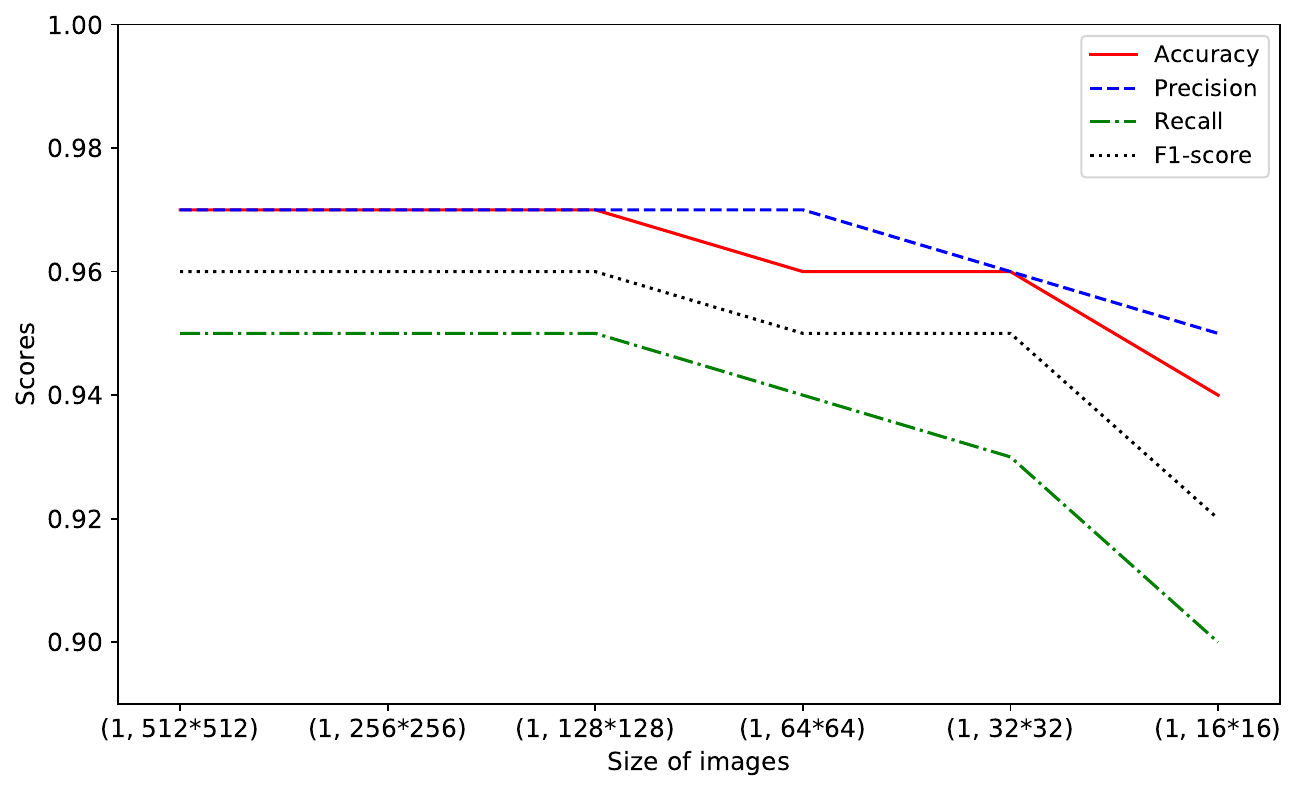}
    \caption{The impact of image-resizing on the performance of \approachname}
    \label{fig:image-size}
\end{figure}

In this section, we study the impact of image-resizing on the effectiveness of our approach.
As we have presented in Section~\ref{sec:approach:images}, we resort to resizing after mapping the bytecode bytes' to pixels in the image.
Since the DEX files can have different sizes, resizing all images to the same size is necessary to feed our neural network.
Resizing implies a loss of information. 
To better assess the impact of image-resizing, we evaluate the performance of \approachname using different image size values. 
Specifically, we repeat the experiment described in Section~\ref{sec:evaluation_rq1} using six sizes for our images: (1, 512x512), (1, 256x256), (1, 128x128), (1, 64x64), (1, 32x32), (1, 16x16). This experiment allows to assess the performance of \approachname over a large range of image sizes, i.e., from $2^8=256$ pixels to $2^{18}=262\,144$ pixels.
We present our results in Figure~\ref{fig:image-size}.

Overall, the size of images is a significant factor in the performance of \approachname. 
Unsurprisingly, the general trend is that the performance decreases as the image size is reduced:
We notice that the values of the four metrics are lower for the three sizes that are smaller than our baseline (1, 128x128), with the worst performance being obtained with the smallest images.
However, increasing the size from (1, 128x128) to either (1, 256x256) or (1, 512x512) does not improve the performance.

\highlight{
\textbf{RQ3 answer:} Image resizing has a significant impact on the effectiveness of our approach.
While downsizing decreases the evaluation scores values by up to 5\%,
increasing the size of our images does not bring significant performance benefits. 
Hence (1, 128x128) seems to be the sweet spot of \approachname.
}

\subsection{\textbf{RQ4:} How does app obfuscation affect the performance of \approachname?}
\label{sec:evaluation:obf}

Malware detectors in the literature are often challenged when presented with obfuscated apps. 
We propose to investigate to what extent \approachname is affected by obfuscation. We consider two scenarios where: (1) the test set includes obfuscated apps; (2) the training set is augmented with obfuscated samples.

\noindent\textbf{Performance on obfuscated apps when \approachname is trained on a dataset of non-obfuscated apps.}
While our non-obfuscated dataset may contain obfuscated samples, we consider it as a normal dataset for training and we assess the performance of \approachname when the detection targets obfuscated apps.
Specifically, we conduct two variant experiments to assess whether \approachname can detect obfuscated malware when: 
(1) It is tested on obfuscated apps that it has seen their non-obfuscated version in the training dataset, 
and (2) It is tested on obfuscated apps that it has NOT seen their non-obfuscated version in the training dataset.
We consider both the non-obfuscated and the obfuscated samples, and we perform our experiments 
using the 10-times Hold-out technique described in Section~\ref{sec:empr:setup}.

For the first experiment, the non-obfuscated dataset is split into 90\% training and 10\% validation.
The test set, which we note \texttt{Test1}, includes all the obfuscated apps.
As for the second experiment, we design it as follows: 
We split the non-obfuscated dataset into 80\% training, 10\% validation, and 10\% test.
The training and the validation apps are used to train and tune \approachname hyper-parameters.
We do not rely on the test images themselves, but we consider their obfuscated versions, which we note \texttt{Test2}.
The average scores of the two experiments are presented in Table~\ref{tab:RQ3_scores}.

We have also evaluated \drebin using the same experimental setup in order to compare with \approachname. 
Its results are also presented in Table~\ref{tab:RQ3_scores}.

\begin{table}[!ht]
	\begin{center}
		\caption{Performance of \approachname \& \drebin on the obfuscated apps}
		\label{tab:RQ3_scores}
            \resizebox{\linewidth}{!}{
		\begin{tabular}{|c | c | c| c| c| c| }
		\hline
			\multicolumn{2}{|c|}{} &  \textbf{Accuracy}  & \textbf{Precision}& \textbf{Recall} & \textbf{F1-score} \\
			\hline
			\multirow{ 2}{*}{\approachname}  & \texttt{Test1} & 0.64  & 0.64 & 0.17 & 0.26\\ 
			& \texttt{Test2} & 0.64  & 0.65 & 0.13 & 0.22\\
			\hline
			\multirow{ 2}{*}{\drebin}  &  \texttt{Test1} & 0.94  & 0.94  & 0.91 & 0.93\\ 
			& \texttt{Test2} & 0.94  & 0.93  & 0.90 & 0.92\\
			\hline
		\end{tabular}}
	\end{center}
\end{table}  

We notice that in the two experiments, \approachname does not perform well on the obfuscated apps detection.
Its scores reported previously in Table~\ref{tab:RQ1_and_2_scores} for malware detection have dropped 
remarkably, especially for the Recall that is decreased by at least 0.78.
The comparison of \drebin's detection scores in Table~\ref{tab:RQ3_scores} and 
Table~\ref{tab:RQ1_and_2_scores} suggests that its effectiveness is slightly decreased on the obfuscated 
apps detection.
The scores reported in Table~\ref{tab:RQ3_scores} are all above 0.9, which demonstrates 
that \drebin's overall performance on the obfuscated apps is still good.
\drebin's results can be explained by the fact that it relies on some features that are not 
affected by the obfuscation process (e.g., requested permissions).

In the rest of this section, we investigate whether augmenting the training dataset with obfuscated samples can help 
\approachname discriminate the obfuscated malware.

\noindent\textbf{Can augmenting the training dataset with obfuscated samples help to discriminate malware?}
With this RQ, we aim to investigate if we can boost \approachname's detection via augmenting the training 
dataset with obfuscated apps.
Specifically, we examine if this data augmentation can improve: 
(1) Obfuscated malware detection, and (2) Malware detection.

We conduct our experiments as follows: We split the non-obfuscated dataset into three subsets: 80\% for the 
training, 10\% for the validation, and 10\% for the test.
We augment the training and the validation subsets with \texttt{X\%} of their obfuscated versions from 
the obfuscated dataset, with \texttt{X = \{25, 50, 75, 100\}}.
As for the test dataset, we evaluate \approachname on both the non-obfuscated images and their obfuscated versions separately.
Specifically, we assess whether augmenting the dataset with obfuscated apps can further enhance:
(1) \approachname's performance on the detection of obfuscated malware reported in Table~\ref{tab:RQ3_scores} (the test set is the obfuscated samples), 
and (2) \approachname's effectiveness reported in Table~\ref{tab:RQ1_and_2_scores} (the test set contains non-obfuscated apps).

Similarly, we evaluate \drebin on the same experimental setup, and we report the average prediction scores 
of both \approachname and \drebin in Table~\ref{tab:RQ4_scores}.

\begin{table}[!ht]
	\begin{center}
		\caption{Performance of \approachname \& \drebin after dataset augmentation}
		\label{tab:RQ4_scores}
		\resizebox{\linewidth}{!}{
		\begin{tabular}{ |c| c| c| c| c| c| }
		\hline
		 &	&  \textbf{Accuracy}  & \textbf{Precision}& \textbf{Recall} & \textbf{F1-score} \\
			\hline
    \multirow{4}{*}{\approachname tested on Obf-apps} &\textbf{25\%} & 0.95  & 0.96 & 0.92 & 0.94\\ 
                    	                              &\textbf{50\%} & 0.96  & 0.97 & 0.92 & 0.95\\
                    		                          &\textbf{75\%} &  0.96 & 0.97 & 0.93 & 0.95\\ 
                    		                          &\textbf{100\%} &  0.96 & 0.97 & 0.94 & 0.95\\
        \hline
		 \multirow{4}{*}{\drebin tested on Obf-apps} &\textbf{25\%} & 0.96  & 0.97 & 0.93 & 0.95\\ 
                    		                         &\textbf{50\%} & 0.96  & 0.97 & 0.94 & 0.95\\
                    		                         &\textbf{75\%} &  0.97 & 0.97 & 0.94 & 0.96\\ 
                    		                         &\textbf{100\%} &  0.97 & 0.97 & 0.94 & 0.96\\
		
        \hline
     \multirow{4}{*}{\approachname tested on non-Obf-apps} &\textbf{25\%} & 0.97  & 0.97 & 0.94 & 0.96\\
                    		                               &\textbf{50\%} & 0.97  & 0.97 & 0.94 & 0.96\\
                    		                              &\textbf{75\%} &  0.97 & 0.97 & 0.94 & 0.96\\ 
                    		                              &\textbf{100\%} & 0.97  & 0.97 & 0.94 & 0.95\\
        \hline
		\multirow{4}{*}{\drebin test on non-Obf-apps} &	\textbf{25\%} & 0.97  & 0.97 & 0.94 & 0.96\\ 
                    		                          &\textbf{50\%} & 0.97  & 0.97 & 0.94 & 0.96\\
                    		                          &\textbf{75\%} & 0.97  & 0.97 & 0.94 & 0.96\\ 
                    		                          &\textbf{100\%} & 0.97  & 0.97 & 0.94 & 0.96\\
        \hline
        
		\end{tabular}}
	\end{center}
\end{table} 

We can see that \approachname detection scores on obfuscated samples increase remarkably when adding obfuscated 
apps to the training dataset.
With 100\% data augmentation, \approachname achieves a detection performance that is comparable to its 
performance on malware detection reported in Table~\ref{tab:RQ1_and_2_scores}.
As for the impact of data augmentation on malware detection, we notice that the detection scores are stable.
These results suggest that data augmentation boosts \approachname detection on obfuscated malware, but it does 
not affect its performance on malware detection (non-obfuscated apps).

As for \drebin, we notice that its performance on the obfuscated apps is also improved, and it is comparable to its performance on malware detection reported in Table~\ref{tab:RQ1_and_2_scores}.
Similarly, its effectiveness on the non-Obfuscated apps after data augmentation is not enhanced, but it is stable.

\highlight{
\textbf{RQ4 answer:} Obfuscation can have a significant impact on the performance of \approachname. When the training set does not include obfuscated apps, the performance on obfuscated samples is significantly reduced. However, when the training set is (even slightly) augmented with obfuscated samples, \approachname can maintain its high detection rate. 
}

\subsection{\textbf{RQ5:} How effective is \approachname in the classification of Android malware families? }
\label{sec:evaluation:rq5}
To better assess the threat of Android malware, security analysts group malicious apps into families~\citep{electronics9060942}.
Each malware family contains apps that exhibit similar malicious behaviour.
Besides distinguishing malware from benign apps, Android malware detectors can also be trained to classify malware apps into families.

In this section, we aim to assess the effectiveness of \approachname in malware family classification.
To this end, we leverage AVCLASS~\citep{sebastian2016avclass} to infer the family labels of our malware samples.
When it is provided with the VirusTotal report of a given malicious app, AVCLASS outputs the most likely family label that can be attributed to this app.
We present in Table~\ref{tab:families_statistics} an overview of the top ten malware families that exist in our dataset:

\begin{table}[!ht]
	\begin{center}
		\caption{An overview of the top ten malware families}
		\label{tab:families_statistics}
		\resizebox{\linewidth}{!}{
		\begin{tabular}{|c |c | c| c| }
		\hline
		      Family name & Nbr of samples & Family name & Nbr of samples \\
		      \hline
			      jiagu          & 35\,623 & hypay  & 104 \\
			      secneo         & 540     & hiddad & 103 \\
			      tencentprotect & 534     & kuguo  & 101 \\
			      dnotua         & 351     & ewind  & 77  \\
			      smsreg         & 288     & wapron & 74  \\
			 
			\hline
		\end{tabular}}
	\end{center}
\end{table} 

We note that AVCLASS could not identify the family label for 37.5\% of our malware apps. These apps have not been included in the evaluation. 
We also discard families that contain only one malware sample.
Consequently, the total number of malware families that are included in our dataset is 82.

Since \approachname's architecture is originally designed for two-class binary classification, we need to adapt it for multi-class family classification.
Specifically, we change the number of neurons in the last layer to 82 neurons, which correspond to the number of family labels.
Moreover, we increase the number of neurons in the first dense layer from 64 to 256 neurons, to ensure that this layer captures sufficient information that can be leveraged by the last dense layer (which now has 82 neurons).

In our experiment, We split each malware family into 80\% training, 10\% validation, and 10\% test.
When the family contains less that ten malware, we put one sample in both the validation and the test, and we use the rest of malware in the training.
We train \approachname using the training samples of all the families and we evaluate its performance on the test samples.
Since malware families in our dataset are highly imbalanced, we report the weighted average of the detection scores.
Specifically, for each family, we multiply its Recall, Precision, and F1-score by the size of the family, and we average these scores over all the families.
We repeat this experiment ten times, and we report the average of the detection scores over the ten experiments in Table~\ref{tab:families_eval}.
We conduct the same experiment using \drebin, and we report its detection scores in the same table.

\begin{table}[!ht]
	\begin{center}
		\caption{Performance of \approachname against \drebin on malware family classification}
		\label{tab:families_eval}
		\begin{tabular}{ |c |c |c |c |c| }
		\hline
			 & \textbf{Accuracy}  & \textbf{Precision}& \textbf{Recall} & \textbf{F1-score} \\
			\hline
			
			\textbf{\approachname} & 0.98 & 0.97 & 0.98 & 0.97  \\ 
			  \hline
			  \textbf{\drebin }   & 0.98 & 0.98 & 0.98 & 0.98 \\ 
			 \hline

		\end{tabular}
	\end{center}
\end{table} 

From table~\ref{tab:families_eval} we observe that \approachname yields high detection scores for malware family classification.
With an F1-score of 0.97, \approachname demonstrates that it can efficiently classify malware families using image representation of the bytecode.
Compared to \drebin, the two malware detectors show similar classification performance.

\highlight{
\textbf{RQ5 answer:} \approachname can classify malware apps into their families with high effectiveness.
When compared to the state-of-the art, \approachname yields comparable classification performance.
}

\subsection{\textbf{RQ6:} What can \approachname tell us about the possibility to localise malicious code?}
\label{sec:evaluation:rq6}
\approachname learns to distinguish malware and benign applications based on (1, 128x128) vector images. 
There is an interesting property of the simple APK representation method used in this work: it is trivial to \emph{mask} parts of the APK after generating the vector image. 

In this section, we propose to leverage this property to investigate the localisation of the parts of the vector image that are important to the prediction.
In other words, we investigate beyond binary classification of whole APKs with two main goals: 1) Assess whether a whole APK is always necessary to perform malware detection, and 2) Assess if some parts are more important than others, and in particular if some parts are a key factor in the predicted classification. 
To this end, we consider two scenarios: 
One where we only \emph{keep} one part of the image (i.e., Section~\ref{rq6_sufficiency}), and one where we only \emph{mask} one part and keep the rest of the image unchanged (i.e., Section~\ref{rq6_neccessity}). 

\subsubsection{Locations that are \emph{sufficient} for the prediction}\label{rq6_sufficiency}

\customdefinition{Definition}{Sufficiency: A part of the image is \emph{sufficient} for the detection if \approachname predicts the malware app as malware when only this part of the image is kept, and the rest is masked}

To identify the parts of the image that are \emph{sufficient} for the prediction, we propose to mask the image vector and keep only a specific part to verify its relevance.
Specifically, we apply a mask on the image by substituting all its pixels with zeros, except for a specific segment where we keep pixels unchanged.
Then, we examine how \approachname predicts the new image vector.
If \approachname still predicts the image as malware (like before masking), it means that the one unmasked segment is sufficient for the prediction of that application.
However, if the predicted label is benign, we can conclude that the pixels of the masked segment contain patterns that contribute to the malware prediction decision. Consequently, a benign prediction suggests that the unmasked segment is not sufficient for the prediction.

Before investigating the sufficiency of the sub-images, we conduct a preliminary experiment that consists of masking all the pixels of an image.
After feeding this masked image to \approachname, it predicts the image as benign. 
This result suggests that the default classification of \approachname, i.e., when it is fed with an empty image, is \emph{benign}, and hence that \approachname will classify a sample as malware only when it detects a pattern associated to maliciousness.

\finding{Finding}{\approachname predicts a fully masked vector image, i.e., an empty image, as benign}

To examine the sufficiency of the sub-images, we propose to split each image into: 2, 4, and 8 parts.
Then, we calculate the percentage of the malware correctly predicted by \approachname when all the pixels are masked except the sub-image under evaluation.
We conduct our experiments using the first training/test splits of the dataset (Cf. Section~\ref{sec:empr:setup}), and we present in Table~\ref{tab:RQ6_suff} the percentage of malware correctly predicted by \approachname when the images are split into 2, 4, and 8 parts.

\begin{table*}[!ht]
	\begin{center}
		\caption{Percentage of malware correctly predicted by \approachname when the sufficiency of the sub-images is evaluated}
		\label{tab:RQ6_suff}
		\scalebox{0.84}{
		\begin{tabular}{  | c| c| c| c| c| c| c| c| c|  }
		\hline
		 	Images split in 2 &  \multicolumn{4}{c|}{$1^{st} 1/2$}  & \multicolumn{4}{c|}{$2^{nd} 1/2$}  \\ \hline
		 	\% of correct malware &  \multicolumn{4}{c|}{95.59} & \multicolumn{4}{c|}{0.26} \\ \hline  \hline 
		 	
		 	Images split in 4 & \multicolumn{2}{c|}{$1^{st} 1/4$}  & \multicolumn{2}{c|}{$2^{nd} 1/4$} & \multicolumn{2}{c|}{$3^{rd} 1/4$}  & \multicolumn{2}{c|}{$4^{th} 1/4$}  \\\hline 
		 	
		 	\% of correct malware &  \multicolumn{2}{c|}{92.56}  & \multicolumn{2}{c|}{0} & \multicolumn{2}{c|}{0}  & \multicolumn{2}{c|}{2.64} \\  \hline \hline
		 	
		 	Images split in 8 & \multicolumn{1}{c|}{$1^{st} 1/8$}  & \multicolumn{1}{c|}{$2^{nd} 1/8$} & \multicolumn{1}{c|}{$3^{rd} 1/8$}  & \multicolumn{1}{c|}{$4^{th} 1/8$} & \multicolumn{1}{c|}{$5^{th} 1/8$}  & \multicolumn{1}{c|}{$6^{th} 1/8$} & \multicolumn{1}{c|}{$7^{th} 1/8$}  & \multicolumn{1}{c|}{$8^{th} 1/8$} \\\hline
		 	
		 	\% of correct malware &  \multicolumn{1}{c|}{85.61}  & \multicolumn{1}{c|}{0} & \multicolumn{1}{c|}{0}  & \multicolumn{1}{c|}{0} &  \multicolumn{1}{c|}{0}  & \multicolumn{1}{c|}{0} & \multicolumn{1}{c|}{0}  & \multicolumn{1}{c|}{15.35} \\  \hline

	\end{tabular}
	}
	\end{center}
\end{table*}

From Table~\ref{tab:RQ6_suff}, we observe that the first half of the images enables the detection of 95.59\% of the malware. 
This part of the images is thus highly sufficient for malware detection. 
As for the second half, Table~\ref{tab:RQ6_suff} shows that it is not sufficient for the detection since it enables to correctly predict only 0.26\% of malware apps.

\finding{Finding}{Not all parts of the image play an equal role in malware detection. The first half of the vector images is largely sufficient to detect malware, while the second half is almost never sufficient. Indeed, \approachname seems to extract the malicious patterns from the beginning of the DEX.}

When splitting the images into 4 parts, the percentage of detected malware is high on the first quarter of the images. As for the three remaining quarters, they do not seem to contain sufficient information for the detection.
The same observation can be made when the images are split into 8 parts. The first one-eighth sub-image is sufficient for the prediction, and the remaining parts seem to have a low sufficiency.
We also observe that the sufficiency of the last part of the images increases when the number of splits increases (i.e., the size of the sub-images decreases). 
Specifically, the percentage of the detected malware has passed from 0.26\% when the images are split into two to 15.35\% when they are split into eight parts.   

\begin{figure}[!ht]
    \centering
    \includegraphics[keepaspectratio=true,width=\linewidth]{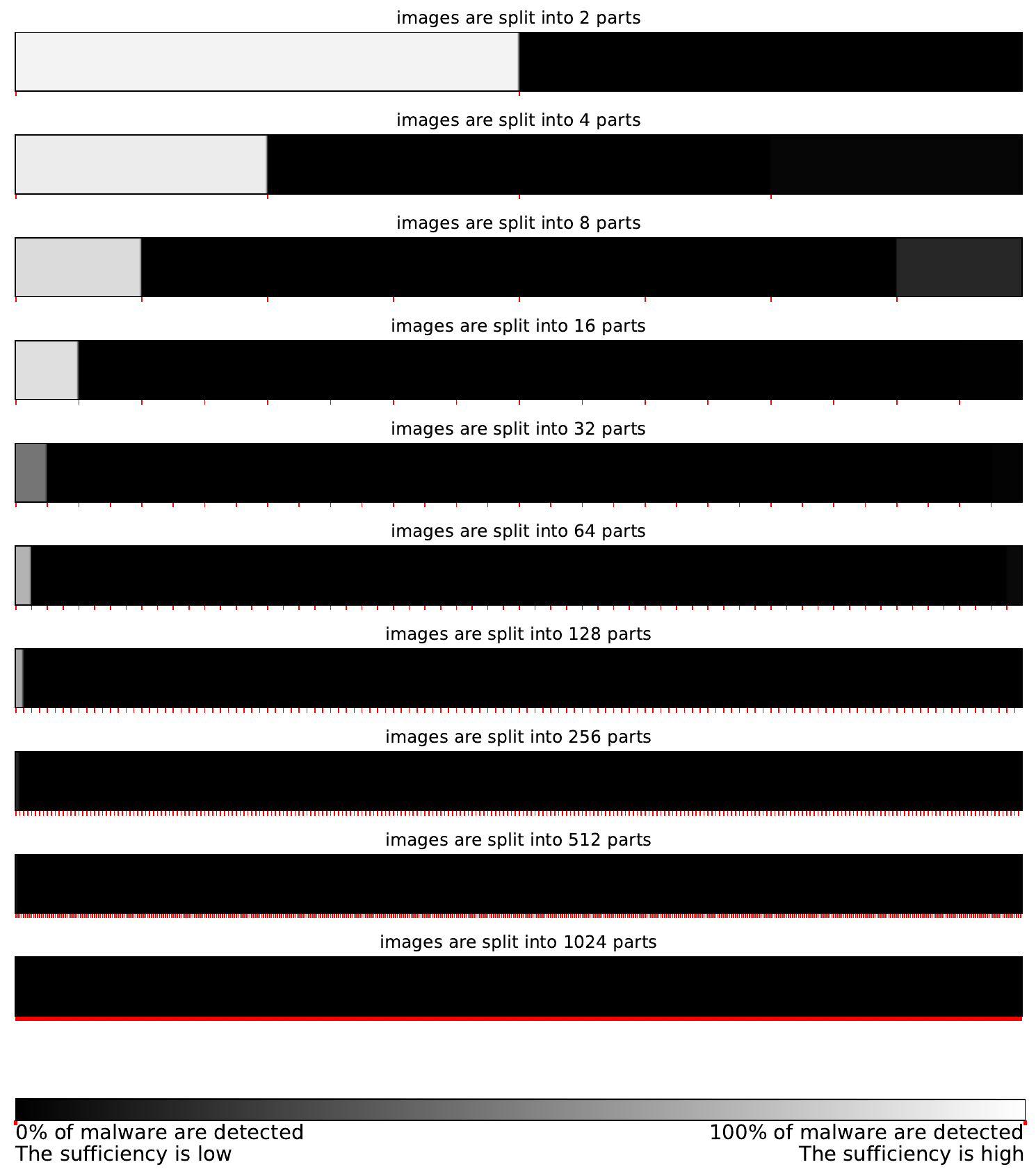}
    \caption{Sufficiency for malware images: High (resp low) sufficiency is represented by white (resp black) colour}
    \label{fig:sufficiency_mal}
\end{figure}

To further confirm these observations, we report in Figure~\ref{fig:sufficiency_mal} the sufficiency of the sub-images when 
the original vector images are split into: 2, 4, 8, 16, 32, 64, 128, 256, 512, and 1024 parts.
The upper plot in Figure~\ref{fig:sufficiency_mal} shows the sufficiency of the first and the second halves of the image vectors (i.e., when the image is split into 2 parts). The sufficiency is represented by the percentage of the apps for which a specific image segment is sufficient for the prediction.
When the sufficiency is high (respectively low), the colour of the image segment is white (respectively black).
Similarly, the second upper plot shows the sufficiency of the first, the second, the third, and the fourth quarters of the vector images (i.e., when the image is split into 4 parts).

From Figure~\ref{fig:sufficiency_mal} we observe that the first half of the image vectors is indeed highly sufficient to identify malware images.
This sufficiency decreases when the number of splits increases. For instance, when the images are split into 256 parts, the sufficiency of the first sub-image seems to be very low compared to the sufficiency of the first sub-images from the upper sub-figures.

\finding{Finding}{The sufficiency of the first pixels in the images generally decreases when their size decreases}

\begin{figure}[!ht]
    \centering
    \includegraphics[width=\linewidth,keepaspectratio=true]{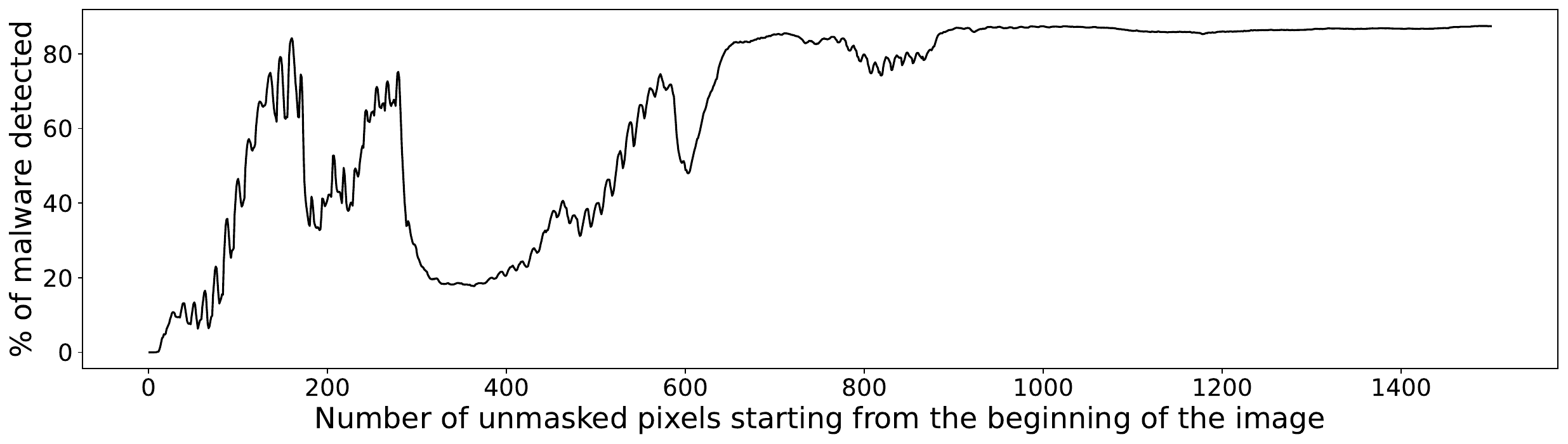}%
    \caption{Sufficiency of the first pixels in details}
    \label{fig:sufficiency_mal_details}
\end{figure}

We also observe that the sufficiency of the first pixels when the images are split into 32 parts is unexpectedly lower than their sufficiency when they are split into 64 parts.
We suspect that the drop of the sufficiency in the first 256 pixels (i.e., the images are split into 32 parts) may be due to an "unlucky" break in the sequence of pixels which has resulted in breaking the sequence of the bytecode that is relevant for malware prediction.
We investigate this behaviour by examining the sufficiency of the first pixels in more details. 
Specifically, we start by keeping only the first pixel in the image and masking all the other pixels.
Then, we unmask gradually one pixel at the time and we evaluate the performance of \approachname. 
We report our results using the first 1500 pixels in Figure~\ref{fig:sufficiency_mal_details}.

The graph in Figure~\ref{fig:sufficiency_mal_details} shows that the sufficiency of the first 890 pixels is highly sensitive to the number of unmasked pixels.
This finding suggests that malicious patterns might be destroyed or noised when some pixels are unmasked.
In other words, providing the classifier with \emph{more} information does not always lead to better classification.
Starting from the first 890 pixels, the sufficiency is stabilised and unmasking the following pixels seems to have no effect on the detection performance.

\finding{Finding}{The sufficiency of the very first pixels is highly sensitive to the number of pixels kept in the image. 
Furthermore, adding more pixels is not always best for the performance}

As for the last pixels in the images, their sufficiency does not seem to increase when the size of the sub-images decreases.
Figure~\ref{fig:sufficiency_mal} shows that the sufficiency of the last sub-images does not continue to increase when the images are split into 16 parts and more.

\finding{Finding}{The last pixels in the vector images are not sufficient to detect malware}

From Figure~\ref{fig:sufficiency_mal}, we also observe that, generally, \approachname needs long sequences of bytecode/pixels to correctly predict malware.

This observation can be verified by calculating the percentage of the total number of malware that is correctly predicted by each of the sub-images separately.
Specifically, we can verify whether the low detection performance in each small sub-image can contribute to a high overall detection performance of \approachname.

For a given split of the images, we calculate the number of malware detected by \approachname using each sub-image separately (e.g., if we split the images in two parts, we sum the number of malware detected with the first part only, with the number of malware detected with the second part only).
Then, we calculate the percentage of the total number of malware detected in all the sub-images after removing the duplicates.
For instance, we can verify whether the total number of malware detected by \approachname when the images are split into 1024 parts is proportional to the total number of malware detected by each of the two halves of the images when they are split in two parts.
We present in Table~\ref{tab:RQ6_total_detected_sufficiency} our results.

\begin{table*}[!ht]
\centering
		\caption{Percentage of total malware correctly predicted by \approachname when evaluating the sufficiency of each  split in the vector images}
		\label{tab:RQ6_total_detected_sufficiency}
		\scalebox{0.85}{
		\begin{tabular}{|c|c|c|c|c|c|c|c|c|c|c|}
		\hline
		 images' splits & $1/2$ & $1/4$ & $1/8$ & $1/16$ & $1/32$ & $1/64$ & $1/128$ & $1/256$ & $1/512$ & $1/1024$ \\ \hline
         \% of detection & 95.68 & 92.76 & 88.15 & 87.69 & 46.65 & 72.54 & 65.91 & 15.49 & 9.41 & 4.03  \\ \hline

	\end{tabular}
	}
\end{table*}

Table~\ref{tab:RQ6_total_detected_sufficiency} shows that when the sub-images are larger, the percentage of the total malware detected by \approachname is higher.
When \approachname is evaluated on sub-images of size 16 (i.e., the images are split into 1024 parts) the percentage of detected malware is very low (i.e., 4.03\%).
This result shows that when the information in the original vector images is divided into small chunks, \approachname is unable to identify patterns relevant for malware prediction decision.

\finding{Finding}{\approachname needs long sequences of bytecode to correctly identify malware}

\subsubsection{Locations that are necessary for the prediction}\label{rq6_neccessity}

\customdefinition{Definition}{Necessity: A part of the image is \emph{necessary} for the detection if \approachname predicts the malware app as benign when this part of the image is masked, and the rest is kept unchanged}

In this second scenario, we propose to investigate the \emph{necessity} of the parts of the images by masking a specific sub-image and keeping the rest of the pixels unchanged.
Specifically, we apply a mask on a specific segment of the image by substituting its pixels with zeros, and we keep the other pixels with their original values.
If \approachname still predicts the image as malware, it means that the masked segment is not necessary for the prediction of that application.
However, if the predicted label is benign, we can conclude that the pixels of the masked segment are necessary for the prediction.
Similarly, we propose to investigate the necessity of the sub-images when the vector images are split into 2, 4, 8 parts.
Table~\ref{tab:RQ6_nece} reports the percentage of malware that is correctly predicted when the necessity of the sub-images is evaluated.

\begin{table*}[!ht]
	\begin{center}
		\caption{Percentage of malware correctly predicted by \approachname when the necessity of the sub-images is evaluated}
		\label{tab:RQ6_nece}
		\scalebox{0.84}{
		\begin{tabular}{  | c| c| c| c| c| c| c| c| c|  }
		\hline
		 	Images split in 2 &  \multicolumn{4}{c|}{$1^{st} 1/2$}  & \multicolumn{4}{c|}{$2^{nd} 1/2$}  \\ \hline
		 	\% of correct malware &  \multicolumn{4}{c|}{0.26} & \multicolumn{4}{c|}{95.59} \\  \hline \hline
		 	
		 	Images split in 4 & \multicolumn{2}{c|}{$1^{st} 1/4$}  & \multicolumn{2}{c|}{$2^{nd} 1/4$} & \multicolumn{2}{c|}{$3^{rd} 1/4$}  & \multicolumn{2}{c|}{$4^{th} 1/4$}  \\\hline 
		 	
		 	\% of correct malware &  \multicolumn{2}{c|}{26.67}  & \multicolumn{2}{c|}{87.18} & \multicolumn{2}{c|}{93.64}  & \multicolumn{2}{c|}{96.46} \\  \hline  \hline
		 	
		 	Images split in 8 & \multicolumn{1}{c|}{$1^{st} 1/8$}  & \multicolumn{1}{c|}{$2^{nd} 1/8$} & \multicolumn{1}{c|}{$3^{rd} 1/8$}  & \multicolumn{1}{c|}{$4^{th} 1/8$} & \multicolumn{1}{c|}{$5^{th} 1/8$}  & \multicolumn{1}{c|}{$6^{th} 1/8$} & \multicolumn{1}{c|}{$7^{th} 1/8$}  & \multicolumn{1}{c|}{$8^{th} 1/8$} \\\hline
		 	
		 	\% of correct malware &  \multicolumn{1}{c|}{52.68}  & \multicolumn{1}{c|}{90.14} & \multicolumn{1}{c|}{93.26}  & \multicolumn{1}{c|}{88.75} &  \multicolumn{1}{c|}{93.22}  & \multicolumn{1}{c|}{95.36} & \multicolumn{1}{c|}{95.57}  & \multicolumn{1}{c|}{94.29} \\  \hline

	\end{tabular}
	}
	\end{center}
\end{table*}

When the images are split into 2 parts, the first half of the images has a high necessity since the percentage of detected malware has dropped to 0.26\% when its pixels are masked.
For the second half, 95.59\% of the malware can be detected by \approachname when its pixels are masked. 
These pixels have thus a low necessity.
We note that when the images are split into two parts, the necessity of the first half is equal to the sufficiency of the second
half and vice-versa.

\finding{Finding}{The first half of the vector images is highly necessary to detect malware}

When the images are split into 4 parts, the necessity of the last two quarters of the images is still low.
For the second quarter, we observe that the percentage of detected malware has increased to 87.18\%, which shows that it has a low necessity.
As for the first quarter, the percentage of detected malware has increased, but its necessity is still high since only 26.67\% of the malware can be identified when its pixels are masked.
The same observation holds for the case when the images are split into eight parts.

\begin{table*}[!htbp]
\centering
		\caption{Percentage of total malware correctly predicted by \approachname when evaluating the necessity of each  split in the vector images}
		\label{tab:RQ6_total_detected_necessity}
		\scalebox{0.85}{
		\begin{tabular}{|c|c|c|c|c|c|c|c|c|c|c|}
		\hline
		images' splits & $1/2$ & $1/4$ & $1/8$ & $1/16$ & $1/32$ & $1/64$ & $1/128$ & $1/256$ & $1/512$ & $1/1024$ \\ \hline
        \% of detection & 95.68 & 97.54 & 97.7 & 97.36 & 97.12 & 97.43 & 98.22 & 97.93 & 97.8 & 97.27  \\ \hline

	\end{tabular}
	}
\end{table*} 

\begin{figure}[!ht]
    \centering
    \includegraphics[keepaspectratio=true,width=\linewidth]{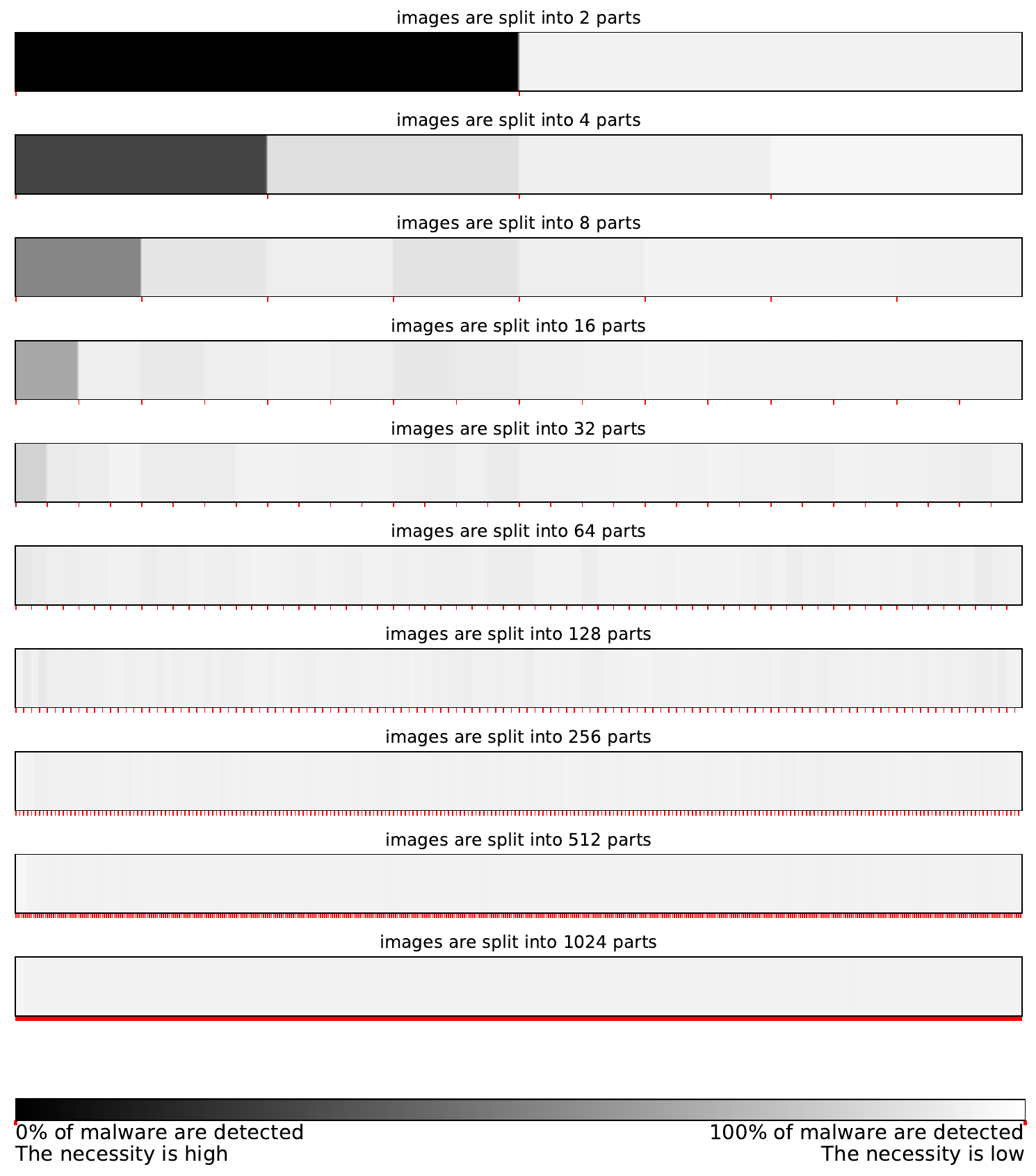}
    \caption{Necessity for malware images: High (resp low) necessity is represented by black (resp white) colour}
    \label{fig:necessity_mal}
\end{figure}

We further present in Figure~\ref{fig:necessity_mal} the necessity of each part of the images when they are split into: 2, 4, 8, 16, 32, 64, 128, 256, 512, and 1024 parts.
We remind that a high percentage of detected malware indicates a low necessity. 
Consequently, when the necessity is high (respectively low), the colour of the image segment is black (respectively white).

The results in Figure~\ref{fig:necessity_mal} show that the first pixels in the images are highly necessary for malware prediction.
This necessity is higher when the size of the sub-images is larger.

\finding{Finding}{The necessity of the first pixels in the images is positively correlated with their size}

Figure~\ref{fig:necessity_mal} also shows that the small segments seem to have a low necessity. This observation is confirmed by Table~\ref{tab:RQ6_total_detected_necessity} that shows the percentage of the total malware detected by \approachname when the necessity of each sub-image is evaluated separately.
For instance, when the sub-images have a size of 16 (i.e., the images are split into 1024 parts), \approachname can detect 97.27\% of the malware.
Consequently, \approachname can maintain its high detection performance in the absence of these small sub-segments.

\finding{Finding}{\approachname's detection performance is not impacted by the absence of a few pixels in the images}

Our experiments have highlighted the sufficiency and the necessity of the first half of the images in detecting malware. 
We aimed to understand the inner-working of \approachname and identify the parts of the images that drive the malware prediction decision.
Consequently, we cannot assume that only the first half of the images is needed for an Android malware detection experiment because that would result in many false positives.
Specifically, the first half of the images has a low sufficiency and necessity in benign apps, which would lead to classifying many benign apps as malware if only the first part of the images is used.
We show in Figure 1 and Figure 2 in the appendix the sufficiency and the necessity of the sub-images that are represented by the percentage of benign apps that are correctly detected as benign by \approachname.

\highlight{
\textbf{RQ6 answer:} Malicious patterns are generally localised at the beginning of the images. Indeed, the first pixels seem to be highly sufficient and necessary for malware prediction. When the first half of the images is split into sub-segments, the sufficiency and the necessity decrease. Moreover, \approachname's performance is not impacted by the absence of a few pixels.
}

\section{Discussion}
\label{sec:discussion}

\subsection{Simple but Effective}
Prior works that propose image-based approaches to Android malware detection build on advanced ``rectangular'' image representations and/or complex network architectures. 
While they achieve high detection rates, the various levels of sophistication in different steps may {\em hide} the intrinsic contributions of the basic underlying ideas. 
With \approachname, we demonstrate that a minimal approach (i.e., with straightforward image generation and a basic CNN architecture) can produce high detection rates in detecting Android malware.

Our experimental comparison against the state-of-the-art detector \drebin further reveals that \approachname is competitive against \drebin. 
In the absence of artefacts to reproduce \rd and \yx's approaches, our comparison focused on the detection scores reported by the authors. 
Note that while \approachname yields similar scores, both approaches involve a certain level of complexity: 
they both rely on 2-dimensional convolution that needs more computational requirements 
than the simple form of convolution leveraged by \approachname.
Moreover, \yx's best architecture includes a high-order feature layer that is added to four extraction units (convolution/pooling layers). 
As for \rd, in addition to the coloured images that require a three-channel representation, it already
leverages a sophisticated convolutional neural network with 42 layers~\citep{inceptionv3}.
Besides, 2-d convolution may not be suitable for image representation of code since pixels from one column of the image are not related. 
Such sophisticated design choices might affect the real capabilities of image-based malware detectors.

\subsection{The Next Frontier in Malware Detection?}

Selecting the best features for malware detection is an open question. 
Generally, every new approach to malware detection comes with its set of features, which are a combination of 
some known features and some novel hand-crafted feature engineering process. 
Manual feature engineering is however challenging and the contributions of various features remain poorly studied in the literature. 

Recent deep learning approaches promise to automate the generation of features for malware detection. 
Nevertheless, many existing approaches still perform some form of feature selection (e.g., APIs, AST, etc.).
Image-based representations of code are therefore an appealing research direction in learning features without a priori selection. 

\approachname's effectiveness suggests that deep learned feature sets could lead to detectors that outperform those created with hand-crafted features.
With \approachname, we use only the information contained in the DEX file, but still we achieve a detection 
performance comparable to the state of the art in the literature. 
This research direction therefore presents a huge potential for further breakthroughs in Android malware detection.
For instance, the detection capability of \approachname can be further boosted using the image representation of 
other files from the Android APKs (e.g., the Manifest file).
We have also revealed that \approachname is not resilient to obfuscation, which calls for investigations into adapted image representations and neural network architectures.
Nevertheless, we have demonstrated that the performance of \approachname is not affected by the time decay, and it also classifies malware families with high effectiveness.
Overall, emerging image-based deep learning approaches to malware detection are promising as the next frontier of research in the field: with the emergence of new variants of malware, automated deep feature learning can overcome prior challenges in the literature for anticipating the engineering of relevant features to curb the spread of malware.

\subsection{Explainability Concerns}

Explainable AI is an increasingly important concern in the research community.
In deep learning based malware detection, the lack of explainability may hamper adoption due to the lack of information that would enable analysts to validate detectors' performance.
Specifically, with \approachname, we wonder: how even a straightforward approach can detect malware with such effectiveness? Are malicious behaviours that easy to distinguish? What do image-based malware detectors actually learn?
The results of our work call for further investigation of the explainability of such approaches.
In our study, we have shown that \approachname relies specifically on the first half of the images to detect malware.
Indeed, the bytecode at the beginning of the DEX file is highly sufficient and necessary for malware detection.
However, our method could not localise specific pieces of malicious code due to the large size of the DEX files.
The resizing of the images has also made it difficult since a small location in the vector images can represent a large sequence of bytecode in the DEX files.
Nevertheless, our experiments have shown which parts of the images are relevant for \approachname predictions.

Following up on the classical case of the wolf and husky detector that turned out to be a snow detector~\citep{lime}, we have concerns as to whether image-based malware detectors learn patterns that humans can interpret as signs of maliciousness.
A more general question that is raised is whether such approaches can eventually help to identify the relevant code that implements the malicious behaviour in an app.
Fine-grained malware localisation and interpretation is indeed important as it is essential for characterising variants and assessing the severity of maliciousness.

\subsection{Threats to validity}
\label{sec:limitations}
Our study and conclusions bear some threats to validity that we attempted to mitigate. 

Threats to {\bf external validity} refer to the generalisability of our findings. Our malware dataset indeed may not be representative of the population of malware in the wild. We minimised this threat by considering a large random sampling from AndroZoo, and by further requiring the consensus of only 2 antivirus engines to decide on maliciousness.
Similarly, our results may have been biased by the proportion of obfuscated samples in our dataset. We have mitigated this threat by performing a study on the impact of obfuscation.

Threats to {\bf internal validity} relate to the implementation issues in our approach. First, \approachname does not consider code outside of the DEX file.
While this is common in current detection approaches that represent apps as images, it remains an important threat to validity due to the possibility to implement malware behaviour outside the main DEX files.
Future studies should investigate apk to image representations that account for all artefacts.
Second, we have relied on third-party tools (e.g., \obfuscapk), which fail on some apps, leading us to discard apps that we were not able to obfuscate or for which we cannot generate images.
This threat is however mitigated by our selection of large dataset for experiments.
Finally, setting the parameters of our experiments may create some threats to validity. For example, since Android apps differ in size, the generated images also have different sizes, which requires to resize all images in order to feed the Neural Network.
The impact of image-resizing on the performance of our approach has been investigated in Section~\ref{sec:evaluation:resizing}.

\section{Related Work}
\label{sec:related_work}
Since the emergence of the first Android malware apps more than ten years ago~\citep{palumbo2017pragmatic}, 
several researchers have dedicated their attention to develop approaches to stop the spread of malware.
Machine Learning and Deep Learning techniques have been extensively leveraged by malware detectors using features extracted either using static, dynamic or hybrid analysis. 
We present these approaches in Section~\ref{sec:related:ml} and Section~\ref{sec:related:dl}. 
We also review related work that leverages the image representation of code for malware detection in 
Section~\ref{sec:related:img}.
Finally, we present malware family classification approaches in Section~\ref{sec:related:fam} and related studies that have contributed to malware code localisation in Section~\ref{sec:related:loc}.

\subsection{Machine Learning-based Android malware detection}\label{sec:related:ml}
Recent studies have been proposed to review and summarise the research advancement in the field of machine learning-based Android malware detection~\citep{9130686, SHARMA2021100373}.
In 2014, \drebin~\citep{drebin} has made a breakthrough in the field by proposing a custom feature set based on static analysis.
\drebin trains a Linear SVM classifier with features extracted from the DEX files and the Manifest file.
Similarly, a large variety of static analysis-based features are engineered and fed to machine learning techniques to detect Android malware (e.g., MaMaDroid~\citep{mamadroid}, RevealDroid~\citep{revealdroid}, DroidMat~\citep{DroidMat}, ANASTASIA~\citep{ANASTASIA}).
Dynamic analysis has also been leveraged to design features for malware detection (e.g., DroidDolphin~\citep{DroidDolphin}, Crowdroid~\citep{Crowdroid}, DroidCat~\citep{droidcat}).
While the above detectors rely either on static or dynamic analysis, some researchers have chosen to rely on features that combine the two techniques (e.g., Androtomist~\citep{kouliaridis2020two}, BRIDEMAID~\citep{BRIDEMAID}, SAMADroid~\citep{SAMADroid}).

All the referenced approaches require feature engineering and pre-processing steps that can significantly increase the complexity of the approach. Our method learns from raw data and extracts features automatically during the training process.

\subsection{Deep Learning-based Android malware detection}\label{sec:related:dl}
Deep Learning techniques have been largely adopted in the development of Android malware detectors.
They are anticipated to detect emerging malware that might escape the detection of the conventional 
classifiers~\citep{7846747}. 
A recent review about Android malware detectors that rely on deep learning networks has been 
proposed~\citep{qiu:2020}. 
MalDozer~\citep{karbab2018maldozer}, a multimodal deep learning-based detector~\citep{kim2018multimodal}, DroidDetector~\citep{7399288}, DL-Droid~\citep{alzaylaee2020dl}, 
Deep4Mal\-Droid~\citep{7814490}, and a deep autoencoder-based approach~\citep{wang2019effective} are examples of 
DL-based detectors that predict malware using a variety of hand-crafted features. 
For instance, MalDozer~\citep{karbab2018maldozer} is a malware detection approach that uses as features the 
sequences of API calls from the DEX file. 
Each API call is mapped to an identifier stored in a specific dictionary.
MalDozer then trains an embedding word model word2vec~\citep{NIPS2013_9aa42b31} to generate the feature 
vectors. 
Again, most of these approaches require a feature engineering step and/or huge computation requirement that is 
not needed by \approachname.

\subsection{Image-based malware detection}\label{sec:related:img}
Different than traditional methods, some approaches have been proposed to detect malware based on the classical ``rectangular'' image-representation of source code.
In 2011, a malware detection approach~\citep{nataraj2011malware} that converts malware binaries into grey images, and extracts texture features using GIST~\citep{oliva2001modeling} has been proposed.
The features are fed to a KNN algorithm for classification purposes. 
Another study~\citep{8626828} has considered the same features for Android malware detection.
Recently, a study~\citep{Yadav2021} has been published about the use of image-based malware analysis with deep learning techniques.

In the Android ecosystem, a malware detection approach~\citep{nver2020} with three types of images extracted from (1) Manifest file, (2) DEX file, and (3) Manifest, DEX and Resource.arsc files has been proposed. 
From each type of image, three global feature vectors and four local feature vectors are created. 
The global feature vectors are concatenated in one feature vector, and the bag of visual words algorithm is used to generate one local feature vector.
The two types of vectors for the three types of images have been used to conduct a set of malware 
classification experiments with six machine learning algorithms. 
Similarly, an Android (and Apple) malware detector~\citep{Mercaldo2020} trained using features extracted from grey-scale binary images has been proposed. 
The method creates the histogram of images based on the intensity of pixels and converts the histogram to 256
features vectors. 
The authors have experimented with different deep learning architectures, and the best results are achieved using a model with ten layers.
R2D2~\citep{R2D2} and \yx~\citep{ding2020android} have proposed malware detectors that are also based on image-learning and they are discussed in detail in Section~\ref{sec:empr:setup}.
None of the above approaches has considered an image representation that preserve the continuity of the bytecode in DEX files.
Moreover, they perform further pre-processing on the images before automatically extracting the features or rely on sophisticated ML or DL models for features extraction and classification.
Our method converts the raw bytecode to ``vector'' images and feeds them directly to a simple 1-dimensional-based CNN model for features extraction and classification.

\subsection{Malware family classification}\label{sec:related:fam}
Several approaches have been proposed to classify Android malware into families.
DroidSieve~\citep{droidsieve} uses static features (e.g., Certificates and Permissions) with Extra Tree algorithm~\citep{geurts2006extremely}.
DroidLegacy~\citep{10.1145/2556464.2556467} relies on API call-based signatures for malware family classification.
Andro-Simnet~\citep{8514216} leverages hybrid features (e.g., API call sequences and Permissions) and Social Network analysis. 
Recently, the image representation of bytecode has been leveraged for malware family classification~\citep{8955840, sun2019android, 10.1145/3426020.3426069}.
For instance, the DEX file's information section has been converted into an image that is used to extract texture and colour features~\citep{8955840}. 
These features have been then fed together with plain text features to the multiple kernel learning classifier.
We note that image-based malware family classification approaches in the literature either rely on advanced architectures, sophisticated images, or (and) perform further image pre-processing.

\subsection{Malicious code localisation}\label{sec:related:loc}
Research on Malicious code localisation, i.e., trying to find what piece of a known malware is actually malicious, has gained significant traction in the Android realm.
Droidetec~\citep{ma2020droidetec} considers the instructions related to the “invoke-” opcodes as natural language sequences and uses the attention mechanism~\citep{vaswani2017attention} to identify features relevant for malicious code localisation. 
The attention layer is integrated into the Bi-LSTM~\citep{hochreiter9ja1} network that is leveraged for malware detection.
MKLDROID~\citep{narayanan2018multi} is a malware detector that considers five views from Android apps and leverages their program representation graphs. This approach uses Multiple Kernel Learning~\citep{gonen2011multiple} SVM to attribute a maliciousness score to each basic block which can trace back the method and class levels.
VizMal~\citep{bacci2018vizmal} builds a trace classifier using the apps' dynamic execution traces. This classifier outputs confidence scores used to generate images that translate the level of maliciousness of the app.

In Image-based malware detection, related approaches~\citep{9307643, 9533803, IADAROLA2021102198, wu2021obfuscation} focus on identifying regions of the square image that contribute to the classification of malware families.
These approaches mainly rely on Grad-CAM~\citep{selvaraju2017grad}, which uses the gradients to produce a heatmap that can be superposed to the input image to highlight important regions for the prediction.
Grad-CAM is more dedicated to visually explaining the prediction of square images.
Since we do not rely on a square image representation of the apps, the use of Grad-CAM on vector images does not seem relevant.

To the best of our knowledge, no image-based malware detector has been used to investigate the localisation of malicious code in the case of binary classification. Instead, they focused on localising regions that discriminate malware families, not investigating regions of interest to discriminate malware against goodware in general.

\section{Conclusion}
\label{sec:conclusion}
We have conducted an investigation to assess the feasibility of leveraging a simple and straightforward approach to malware detection based on image representation of code. \approachname implements a 1-dimensional convolution with two extraction units (Convolution/Pooling layers) for the neural network architecture.
The evaluation of \approachname on a large dataset of malware and benign Android apps demonstrates that it achieves a high detection rate.
We have also showed that our approach is robust against time decay, and studied the impact of image-resizing and obfuscation on its performance.
Moreover, we have demonstrated that \approachname is highly effective in classifying malware families, and we have investigated the localisation of parts of vector images that are relevant to malware detection.

We have also compared \approachname against prior work on Android malware detection. 
Our results demonstrate that \approachname performs similarly to the state of the art \drebin and two image-based detectors that consider more sophisticated network architectures.
The high performance of \approachname suggests that image-based Android malware detectors are indeed promising. We expect our work to serve as a foundation baseline for further developing the research roadmap of image-based malware detectors.
We release the dataset of apps and the extracted images to the community.
We also make \approachname source code publicly available to enable the reproducibility of our results
and enable other researchers to build on our work to further develop this research direction.

\section{Data Availability}
\label{data_availability}

The datasets and artefacts used in the present study are available in our repository:
\url{https://github.com/Trustworthy-Software/DexRay}




\section{Acknowledgements} 

This work was partially supported 
(a) by the  Fonds National de la Recherche (FNR), Luxembourg, under project Reprocess \\ C21/IS/16344458, 
(b) by the University of Luxembourg under the HitDroid grant,
and (c) by the Luxembourg Ministry of Foreign and European Affairs through their Digital4Development (D4D) portfolio under project LuxWAyS.

\section{Conflict of interest}
The authors declare that they have no conflict of interest.

\bibliographystyle{elsarticle-harv}
\bibliography{biblio}

\newpage



\section{Appendix}%
\label{sec:appendix}%
\begin{figure}[htbp]
    \centering
    \includegraphics[width=0.95\linewidth]{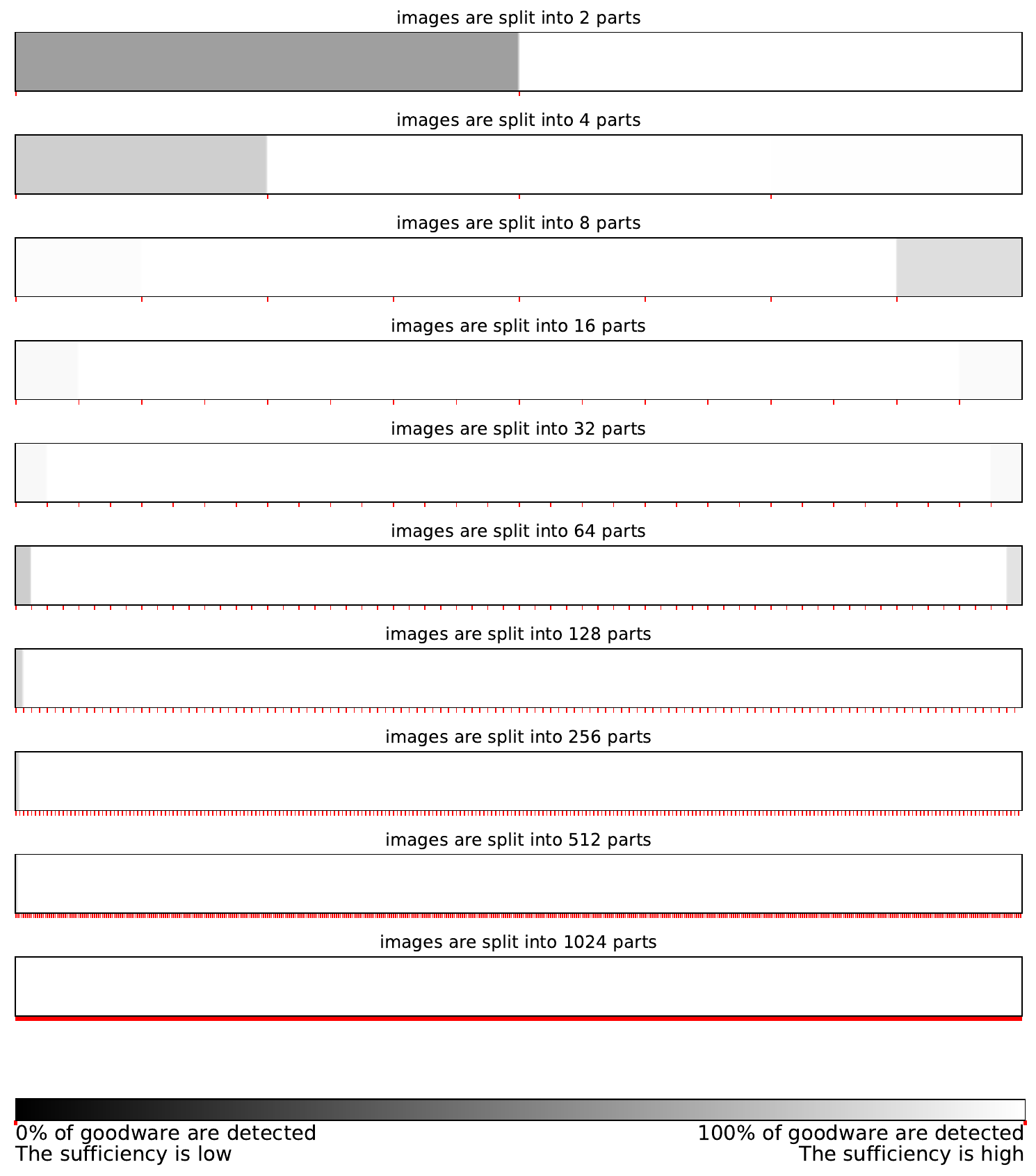}
    \caption{Sufficiency for benign images: High (resp low) sufficiency is represented by black (resp white) colour}\label{fig:sufficiency_good}
\end{figure}

\begin{figure}[h!]
    \centering
    \includegraphics[width=0.95\linewidth]{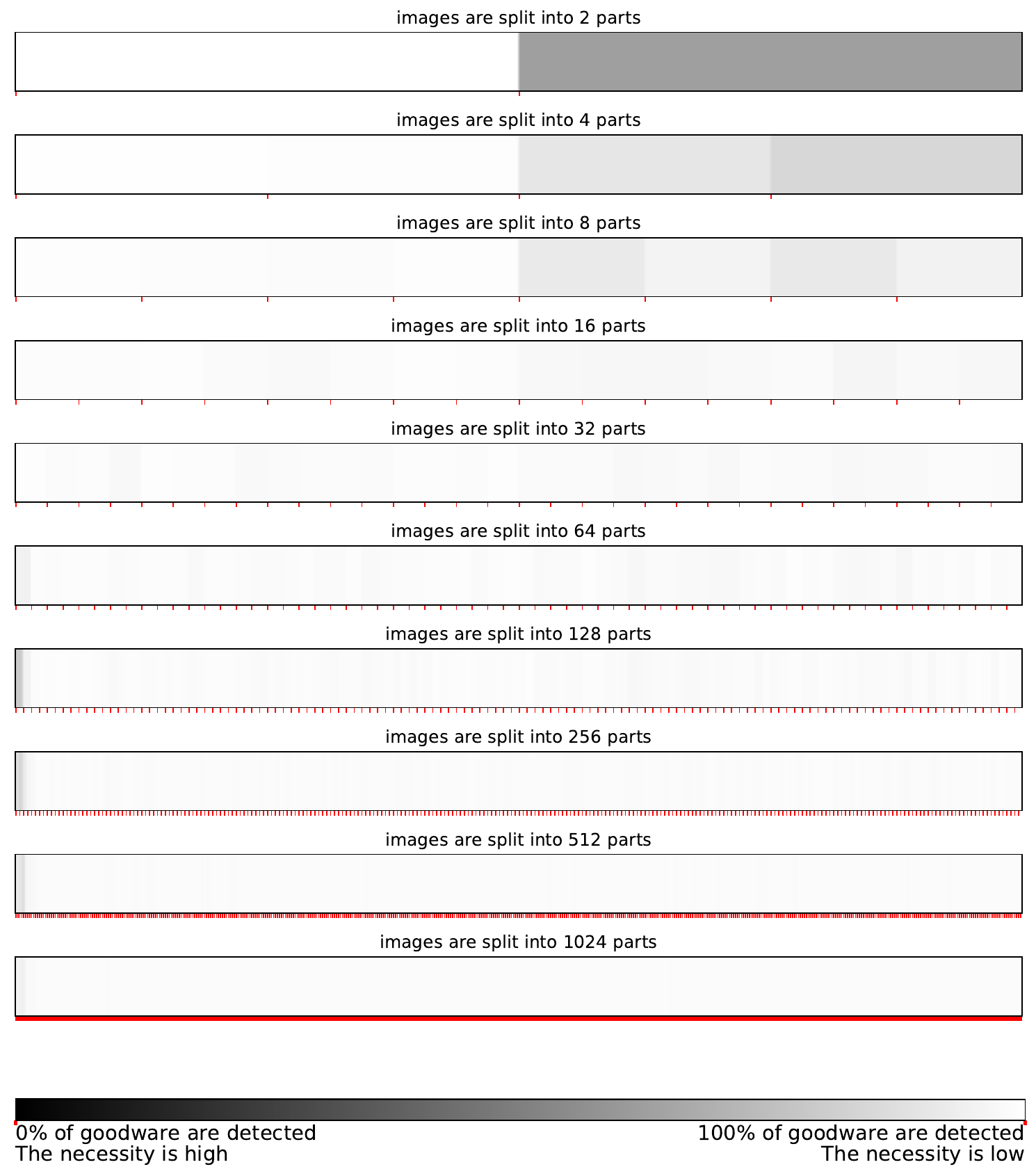}
    \caption{Necessity for benign images: High (resp low) necessity is represented by black (resp white) colour}\label{fig:necessity_good}
\end{figure}

\FloatBarrier
\newpage


\end{document}